\RequirePackage{fix-cm}
\documentclass[smallextended,numbook]{svjour3}       
\smartqed  
\usepackage{graphicx}
\usepackage{amsmath,amssymb}
\usepackage{physics}
\usepackage{upgreek}
\usepackage{color}
\begin{document}

\newcommand{\comment}[1]{{\textcolor{red}{#1}}}
\newcommand{\changed}[1]{{\textcolor{blue}{#1}}}

\title{Influence of weak electromagnetic fields on charged particle ISCOs
}


\author{Jan P. Hackstein \and Eva Hackmann}


\institute{J. P. Hackstein \at
              Center of Applied Space Technology and Microgravity (ZARM), University of Bremen, Am Fallturm 2, 28359 Bremen, Germany \\
              \email{ja\_ha@uni-bremen.de}           
           \and
           E. Hackmann \at ZARM, University of Bremen\\
           \email{eva.hackmann@zarm.uni-bremen.de}
}

\date{Received: date / Accepted: date}

\maketitle

\begin{abstract}
Astrophysical black holes are often embedded into electromagnetic fields, that can usually be treated as test fields not influencing the spacetime geometry. Here we analyse the innermost stable circular orbit (ISCO) of charged particles moving around a Schwarzschild black hole in the presence of a radial electric test field and an asymptotically uniform magnetic test field. We discuss the structure of the in general four ISCO solutions for different magnitudes of the electric and the magnetic field's strength. In particular, we find that the nonexistence of stable circular orbits of particles with equal sign of charge as the black hole for sufficiently strong electric fields can be canceled by a sufficiently strong magnetic field. In this situation, we find that ISCOs made of static particles will emerge.  

\end{abstract}

\section{Introduction}
\label{intro}

According to the no-hair theorem \cite{ChruscielLRR}, an isolated black hole is a rather simple object that can be described by only three parameters\footnote{We could add the magnetic monopole charge as a fourth parameter.}: mass, rotation, and electric charge. In astrophysics, the electric charge is usually neglected: As the electromagnetic interaction is much stronger than the gravitational interaction, the charged black hole will selectively accrete oppositely charged particles from its environment, and will decrease its charge to mass ratio to tiny values\footnote{We use geometrised units $\tilde{Q}= \frac{Q_{\rm SI}\sqrt{G}}{\sqrt{4\pi\epsilon_0}c^2}$ and $\tilde{M}=GM/c^2$, where $Q_{\rm SI}$ is the charge of the black hole, $M$ is the mass of the black hole, $G$ is the gravitational constant, $c$ is the speed of light, and $\epsilon_0$ is the electric constant, all in SI units.}  $\tilde{Q} < 10^{-18} \tilde{M}$ on a short timescale \cite{Eardley1975}. Even if the black hole is completely isolated and cannot accrete from its environment, due to pair production it will reduce its charge to mass ratio quickly to about $\tilde{Q}/\tilde{M} < 10^{-5} M_{\rm BH}/M_{\rm Sun}$ \cite{Eardley1975}. 

The influence of the charge parameter on the curvature of spacetime can for instance be estimated by considering the Kretschmann scalar $K$ of a Reissner-Nordstr\"{o}m black hole \cite{Henry2000},
\begin{align}
K = R^{\mu\nu\rho\sigma} R_{\mu\nu\rho\sigma} = \frac{48\tilde{M}^2}{r^6} - \frac{96\tilde{M}\tilde{Q}^2}{r^7} + \frac{56 \tilde{Q}^4}{r^8}\,.
\end{align}
We see that on horizon scales the influence of $\tilde{Q}$ is comparable to the influence of $\tilde{M}$ if $\tilde{Q} \sim \tilde{M}^2$. Therefore, we can safely neglect the influence of a tiny charge on the curvature of spacetime. However, again as the electromagnetic interaction is much stronger than the gravitational interaction, we should separately consider the effect of the electric field of the black hole on charged particles. The equations of motion of charged particles with specific charge\footnote{In geometrised units  $q=\frac{q_{\rm SI}}{\sqrt{4\pi\epsilon_0 G}m}$, where $m$ is the mass of the particle that is assumed to be much smaller than the mass of the black hole.} $q$ moving around a black hole with charge $\tilde{Q}$ contain the product $q\tilde{Q}/\tilde{M}$ \cite{Grunau2011}. For a free electron (proton), $q$ is very large, of the order $10^{21}$ ($10^{18}$) \cite{Schroven2017}. Therefore, the charge product $q\tilde{Q}/\tilde{M}$ will in general not be small. This order of magnitude estimate suggests to take the charge of the black hole into account as a test field, that does not influence the spacetime geometry.

Recently, charged black holes experienced a renewed interest in the astrophysical literature. For instance, Nathanail et al. \cite{Nathanail2017} studied the likely scenario of the collapse of a rotating neutron star with a magnetosphere and an initial net electric charge, and found that the newly formed black hole has a Kerr-Newman geometry, whose rather large charge to mass ratio $\tilde{Q} \approx 10^{-4} \tilde{M}$ will probably quickly reduce to tiny values according to the processes discussed above. Zhang \cite{Zhang2016} discussed the possibility that the merger of a highly charged ($\tilde{Q}\sim 10^{-9} \tilde{M}$) black hole and a neutral black hole may create a Fast Radio Burst \cite{Lorimer2007,Thornton2013}. Also in the context of Fast Radio Bursts, Punsley and Bini \cite{Punsly2016} discussed the sudden discharging of a Kerr-Newman black hole, and Liu et al. \cite{Liu2016} the collapse of a magnetosphere of a charged black hole. Levin et al. \cite{Levin2018} argued that the binary merger of a charged black hole and a neutron star may create an electromagnetic counterpart to the gravitational wave observation. Recently, Zaja{\v{c}}ek et al. \cite{Zajacek2018} estimated the electric charge of Sagittarius A*, the central supermassive black hole of the Milky Way, from the observation of bremsstrahlung. They found $Q_{\rm SI} \lesssim 3 \times 10^8 \rm C$ or $\tilde{Q} \lesssim 4 \times 10^{-19} \tilde{M}$.

In addition to possessing a tiny electric charge, astrophysical black holes are usually embedded into external electromagnetic fields. In particular, a surrounding accretion disk can produce a magnetic field. Other possibilities are magnetic fields originating from before the gravitational collapse, or completely external fields like the galactic magnetic field or fields from a nearby neutron star/magnetar. Magnetic fields are also believed to play a major role in the creation of jets \cite{Blandford1977}. Note that a rotating, initially neutral black hole will drag along the magnetic field, thereby enabling selective accretion of charges, resulting in a black hole with a stable net electric charge, see for instance \cite{Wald74,King1975,Ruffini1975,Gibbons1976,Aliev1989,Lee2001}. The characteristic scales of magnetic fields near stellar mass and supermassive black holes were estimated in \cite{Piotrovich2010,Frolov2010}, and found to be many orders of magnitudes too small to have a noticeable effect on the spacetime geometry. However, similar to the case of an electric charge discussed above, the large charge to mass ratio of free electrons and protons will compensate the small magnetic field strength and lead to non-negligible Lorentz force effects \cite{Frolov2010}. 

Summarised, the effects due to electric and magnetic fields in the vicinity of a black hole should not a priori be completely neglected, but carefully studied for their influence on charged matter motion. In this paper, we consider a Schwarzschild black hole endowed with an electric charge and embedded into an external magnetic field. Both the electric and the magnetic field will be treated as test fields, meaning that we neglect their effects onto the spacetime geometry. For the magnetic field we choose a particularly simple model, that is, an asymptotically uniform field as discussed by Wald \cite{Wald74}. This kind of magnetic field can be considered as an approximation to the realistic fields discussed above, for instance if the accretion disk is much larger than the black hole, or a rather far away magnetar. Note that for a vanishing electric field this scenario reduces to the one discussed in \cite{Frolov2010}.

A particularly interesting feature of particle motion is the innermost stable circular orbit (ISCO), also called marginally stable orbit. It is a distinctive feature of General Relativity without Newtonian analogue and marks the transition from a region where stable circular orbits are possible to a region close to the black hole where particles will eventually plunge into the central black hole. In the context of accretion disks, it approximately marks the inner edge of the Shakura-Sunyaev geometrically thin disk model \cite{Shakura1973,Abramowicz2013LRR} and the center of the "polish doughnut" geometrically thick accretion disk model with constant angular momentum \cite{Abramowicz2013LRR,Abramowicz1978}. These are regularly used as initial conditions for accretion disk simulations and are therefore also related to the observations by the Event Horizon Telescope \cite{EHT2019_I}. For the future gravitational wave observations from extreme mass ratio inspirals with the LISA mission \cite{Amaro-Seoane2012}, the ISCO marks the onset of the plunge region. In principle the ISCO can also be used to extract information about the rotation and/or the charge of the black hole \cite{Zajacek2018}.

The structure of the paper is as follows. In the next section we introduce the spacetime and the electromagnetic fields in our setup and derive the equations of motion for charged particles. The innermost stable circular orbit is introduced and analysed in section 3. There we first analyse some general features and discuss the limiting case of a vanishing magnetic field. Note that the other limit of a vanishing electric charge was already covered in \cite{Frolov2010}, and we reproduce the general characteristics when considering a very small electric charge in subsection \ref{Q1}. We then further split our discussion into two parts, as $q\tilde{Q}=\tilde{M}$ marks an important transition, discussed before in the context of a Reissner-Nordström black hole in \cite{Pugliese2017}. The paper closes with a summary and discussion.

\section{Metric and electromagnetic fields}
\label{bfields}

The Schwarzschild metric is given by
\begin{equation}
g=-\left(1-\frac{r_\text{s}}{r}\right)c^2\dd t^2+\frac{\dd r^2}{1-\frac{r_\text{s}}{r}}+r^2(\dd \vartheta^2+\sin^2(\vartheta) \dd \varphi^2),
\label{metric}
\end{equation}
where $r_\text{s}=\frac{2GM}{c^2}$ is the Schwarzschild radius. Since the metric is static and spherically symmetric, the two Killing vector fields are $\xi_\text{(t)}=\partial_\text{t}$ and $\xi_{(\upvarphi)}=\partial_\upvarphi$.

A test particle with specific charge $q$ moves according to the equation
\begin{equation}
\dot{u}^\mu=(\ddot{x}^\mu+\Gamma^\mu_{\;\;\nu\rho}\dot{x}^\nu\dot{x}^\rho)=q F^\mu_{\;\;\nu} u^\nu,
\label{Geogl}
\end{equation}
where the right hand side describes the electromagnetic force acting on the test particle, given by the electromagnetic tensor $F^\mu_{\;\;\nu}=(\partial^\mu A_\nu - \partial_\nu A^\mu)$ and the particle's four velocity $u^\nu$ with $u^\nu u_\nu=-1$. An overdot denotes the derivative $\dv{}{\tau}$ with respect to the particle's proper time $\tau$.

To describe the effect of an electromagnetic field on an orbiting test particle, the following four potential was chosen \cite{Wald74}
\begin{equation}
A^\mu=\frac{B}{2}\xi_{(\upvarphi)}^\mu - \frac{Q}{2}\xi_\text{(t)}^\mu,
\label{Pot}
\end{equation}
corresponding to an asymptotically uniform magnetic test field with magnetic field strength $B$ parallel to the z-axis and a radial electric test field of charge $Q$ with its source located in the center of the considered black hole. Without loss of generality, we choose $B>0$ and $Q>0$.

For the purpose of presenting all equations in a concise manner, it is convenient to normalise all parameters with respect to an effective mass $\tilde{M}$, resulting in dimensionless quantities
\begin{align}
	\tilde{M}=\frac{GM}{c^2}, \qq{} Q^2=\frac{Q_\text{SI}^2 G}{4\pi \varepsilon_0 c^4 \tilde{M}^2}, \qq{} B^2=\frac{4\pi\varepsilon_0 G \tilde{M}^2}{c^2}B_\text{SI}^2,\\
	q^2=\frac{q_\text{SI}^2}{4\pi\varepsilon_0 G m^2}, \qq{} r=\frac{r_\text{SI}}{\tilde{M}},\qq{}  \tau=\frac{c}{\tilde{M}}\tau_\text{SI}, \qq{} t=\frac{c}{\tilde{M}}t_\text{SI},
\end{align}
where parameters in geometric units are marked with a tilde, while dimensionless parameters are left unaltered. The effective energy and effective angular momentum are denoted by $E=\frac{E_\text{SI}}{m c^2}$ and $L=\frac{L_\text{SI}}{m \tilde{M} c}$, respectively.

To simplify the following calculations, the test particle's movement was restricted to the equatorial plane ($\vartheta=\frac{\pi}{2}$). An equation of motion was consequently calculated for the radial as well as the angular and the time coordinate with the Lagrangian formalism, 
\begin{align}
	\dot{r}^2 & = \left( 1-\frac{2}{r}\right)^2 \dot{t}^2 - \left( 1-\frac{2}{r}\right) (1+r^2\dot{\varphi}^2)\nonumber\\
	& =\qty\bigg(E+\mathcal{Q}\qty\bigg(1-\frac{2}{r}))^2-\qty\bigg(1-\frac{2}{r})\qty\bigg(1+\frac{(L-\mathcal{B}r^2)^2}{r^2})=:\mathcal{U},
\label{rdot}
\end{align}
\begin{equation}
	\dot{t} = \frac{E}{1-\frac{2}{r}}+\mathcal{Q}, \qq{} \dot{\varphi} = \frac{L}{r^2}-\mathcal{B},
	\label{tphidot}
\end{equation}
where $\mathcal{U}$ denotes the effective potential, $\mathcal{B}=\frac{qB}{2}$ the effective magnetic field strength and $\mathcal{Q}=\frac{qQ}{2}$ the effective charge. Since the particle's electric charge $q$ will be kept constant in this paper, changing the electric or magnetic field is proportional to changing $\mathcal{Q}$ or $\mathcal{B}$. The coupling of the particle's energy $E$ with the test field's electric charge $Q$ is visible in equations \eqref{rdot} and \eqref{tphidot}, whereas its angular momentum $L$ couples with the test field's magnetic field strength $B$. Note that according to equation \eqref{tphidot} the sign of $E$ is not necessarily identical to the sign of $\dot{t}$, and also the signs of $\dot{\varphi}$ and $L$ may in general be different. The equation of motion \eqref{rdot} is invariant under the transformation $(L,\mathcal{B}) \to (-L,-\mathcal{B})$ as well as $(E,\mathcal{Q}) \to (-E,-\mathcal{Q})$. We may therefore choose w.l.o.g. $\mathcal{B}\geq0$ and $\mathcal{Q}\geq0$. In the limiting case of a vanishing electric charge, these equations yield the results discussed in \cite{Frolov2010} for a magnetic test field.

\section{The Innermost Stable Circular Orbit}
\label{ISCO}

The imposed conditions on the effective potential $\mathcal{U}$ for the existence of an ISCO are
\begin{equation}
\mathcal{U}(r)=0\,, \quad \dv{\mathcal{U}}{r}=0\,, \quad \dv[2]{\mathcal{U}}{r}=0\,.
\label{con}
\end{equation}
Substituting the equation of motion \eqref{rdot} into $\dv{\mathcal{U}}{r}=0$ and $\dv[2]{\mathcal{U}}{r}=0$, one can determine the following expressions for the particle's angular momentum $L$ and energy $E$, respectively
\begin{equation}
L_{\upalpha,\upbeta}=\pm \frac{C}{r-6},
\label{Leq}
\end{equation}
\begin{equation}
E_{\upalpha,\upbeta}=\frac{\mathcal{B}^2 (4r^5-18r^4+12r^3)+(1-2\mathcal{Q}^2)r^2+D_{\mp}r-12\mathcal{Q}^2}{2r(r-6)\mathcal{Q}},
\label{Eeq}
\end{equation}
with $C=\sqrt{-(r-6)r(3\mathcal{B}^2 r^4-2\mathcal{B}^2 r^3-4\mathcal{Q}^2)}$ and $D_{\mp}=(-6 \mp 2C\mathcal{B}+12\mathcal{Q}^2)$. Using the first condition of \eqref{con} to calculate the radial component $r$ yields no simple analytical expression. To depict the electromagnetic field's influence on the ISCO's behaviour, the resulting equation was instead solved numerically on a fixed interval of the effective magnetic field strength $\mathcal{B}$ with constant electric charge $Q$. To furthermore gain insight on how the field's electric charge changes the particle's movement, different values for $Q$ were tested, resulting in different figures to be compared. As shown in \cite{Pugliese2017} in the context of a Reissner-Nordstr\"om spacetime, $qQ=1$ or $\mathcal{Q}=\frac12$ presents a limiting case. Therefore, in this paper, the three cases $\mathcal{Q}\ll 1$, $\mathcal{Q}<\frac{1}{2}$ and $\mathcal{Q}\geq \frac{1}{2}$ were examined separately. The results were subsequently interpreted to determine the electromagnetic field's impact on the test particle.

\subsection{General characterisation}
\noindent Inserting either $L_{\upalpha}$ and $E_{\upalpha}$ or $L_{\upbeta}$ and $E_{\upbeta}$ into the first condition of \eqref{con} yields two radii $r$, respectively. Thus one can determine four different solutions $r_1$, $r_2$, $r_3$, $r_4$ for $\mathcal{B}\neq 0$ and $\mathcal{Q}\neq 0$, where $r_{1,2}$ correspond to $(E_\upalpha,L_\upalpha)$ and $r_{3,4}$ to $(E_\upbeta,L_\upbeta)$. An explanation for the appearance of four solutions in the presence of an electromagnetic field can be given when considering the influence of the electromagnetic force $F^\text{EM}$ on the test particle.

Each $r_i$ (with $i \in \{1, ...,4\}$) corresponds to an angular momentum $L_i$ and an energy $E_i$ given by equations \eqref{Leq} and \eqref{Eeq}. Because of the magnetic field's alignment with the z-axis and the particle movement's restriction to the equatorial plane perpendicular to the z-axis, only the two possibilities $\vec{L}_i\uparrow\uparrow \vec{B}$ and $\vec{L}_i\uparrow\downarrow \vec{B}$ remain. Classically, this determines the direction of the particle's momentum $\vec{p}_i$, either resulting in an orbit with $d\varphi/dt>0$, what we call here a direct orbit, or $d\varphi/dt<0$ (indirect orbit). 
The induced Lorentz force $F^\text{L}$ consequently points radially outwards (repelling) or inwards (attracting), classically entirely depending on the particle's angular momentum. The Coulomb force $F^\text{C}$ caused by the electromagnetic field additionally either attracts or repels the test particle depending on the sign of its charge $q$.

The right hand side of equation \eqref{Geogl} describes the relativistic electromagnetic force on the particle compared to the classical case described above. It can be checked that only the $r-$component becomes non-zero for conditions \eqref{con}, resulting in a purely radial Lorentz and Coulomb force. The right-hand side of equation \eqref{Geogl} for $\mu=r$ reads
\begin{equation}
q F^\text{r}_{\;\;\nu} u^\nu=\left(1-\frac{2}{r}\right)\left(\frac{\mathcal{Q}\dot{t}}{r^2}+r\mathcal{B}\dot{\varphi}\right).
\label{relF}
\end{equation}
Since the left bracket will be non-negative in the exterior Schwarzschild region, the forces' sign depends on the right bracket, where $\dot{t}$ and $\dot{\varphi}$ are given by equations \eqref{tphidot}. As we chose w.l.o.g. $\mathcal{Q}>0$ and $\mathcal{B}>0$, the sign of the Coulomb force is directly given by the sign of $\dot{t}$, and the sign of the Lorentz force by the sign of $\dot{\varphi}$. As we physically of course move forward in time, we can interpret the mathematical result $\dot{t}<0$ by simultaneously switching the signs of $E$ and $\mathcal{Q}$, which leaves the equation of motion \eqref{rdot} and the force equation \eqref{relF} invariant. Therefore, $\dot{t}<0$ corresponds physically to a particle that has actually $\dot{t}>0$ and $\mathcal{Q}<0$. Similarly, as we chose $B>0$ in the beginning, the motion of a negatively charged particle $\mathcal{B}<0$ is equivalent to the motion of a positively charged particle $\mathcal{B}>0$ with opposite sign of $\dot{\varphi}$. Summarised, the four solutions $r_i$, $i \in \{1, ...,4\}$, are physically related to the four combinations $\mathcal{Q}\dot{t} \gtrless 0$, $\mathcal{B}\dot{\varphi} \gtrless 0$.

\subsection{Electromagnetic field with $\mathcal{B}=0$}\label{Qx}

\begin{figure}
	\centering
	\includegraphics[width=0.6\textwidth]{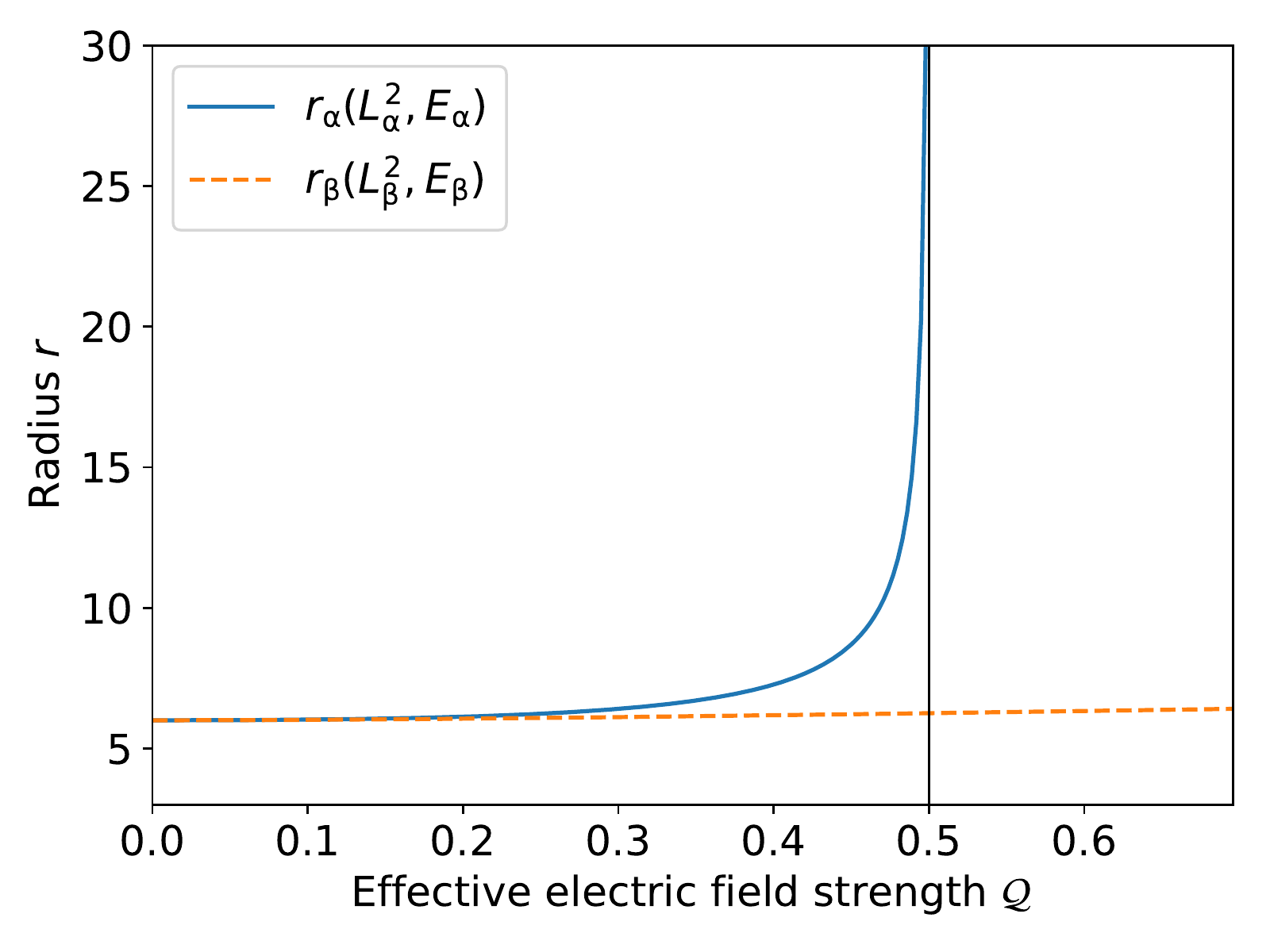}
	\caption{Radii $r_{\upalpha,\upbeta}$ of the ISCO for charged test particles in an electric field as a function of the effective electric field strength $\mathcal{Q}$. Each radius corresponds to orbits of either a positively ($r_{\upalpha}$) or negatively ($r_{\upbeta}$) charged particle. The black, vertical line emphasizes $r_{\upalpha}$ diverging at $\mathcal{Q}=\frac{1}{2}$. As a consequence, only negatively charged particles form an ISCO for a sufficiently strong electric field strength.}
	\label{fig:7}
\end{figure}

As the impact of a purely magnetic test field in the Schwarzschild spacetime was already examined in \cite{Frolov2010}, we initially neglect  $\mathcal{B}$ to investigate the behaviour of a test particle in an electric field for later reference. This situation was visualized in figure \ref{fig:7} by choosing $\mathcal{B}=0$ and plotting the radii $r_i$ on a fixed interval $\mathcal{Q}=[0,0.7]$. 

As can be seen, two solutions $r_{\upalpha,\upbeta}$ exist in the region $0<\mathcal{Q}<0.5$. Since equation \eqref{rdot} only depends on $L^2$ for $\mathcal{B}=0$, solving one of the conditions \eqref{con} for $L^2$ is possible, meaning $r$ does not depend on the angular momentum's sign. In this illustration, $r_\upalpha$ corresponds to a positively and $r_\upbeta$ to a negatively charged particle. For $\mathcal{Q}=0$, both solutions converge to $r=6$, the ISCO of neutral particles in a Schwarzschild spacetime. For higher electric field strengths, the distance between $r_\upalpha$ and $r_\upbeta$ increases until $r_\upalpha$ diverges at $\mathcal{Q}=0.5$. The electromagnetic force given by equation \eqref{relF} for the diverging radii was examined in this limit. Even though $\mathcal{B}=0$ already negates the Lorentz force, the angular velocity $\dot{\varphi}$ converges to zero, corresponding to a static particle at a fixed coordinate angle $\varphi$. The Coulomb force additionally vanishes for diverging radii, because $\dot{t}$ is finite in this limit. This results in $F^\text{EM}_{\upalpha}=0$. The limit therefore represents a static particle at radial infinity without an electromagnetic force acting on it. From this, it seems reasonable that no solution exists for an ISCO of positively charged particles from this point onward in figure \ref{fig:7}. An analogous behaviour was already studied for a charged test particle in the gravitational field of a charged Reissner-Nordstr\"{o}m black hole \cite{Pugliese2011}. For $\mathcal{Q}>\frac{1}{2}$, only $r_\upbeta$ corresponding to a negatively charged particle exists.

\subsection{Electromagnetic field with $\mathcal{Q}\ll 1$}
\label{Q1}
\begin{figure}
	\centering
	\includegraphics[width=0.495\textwidth]{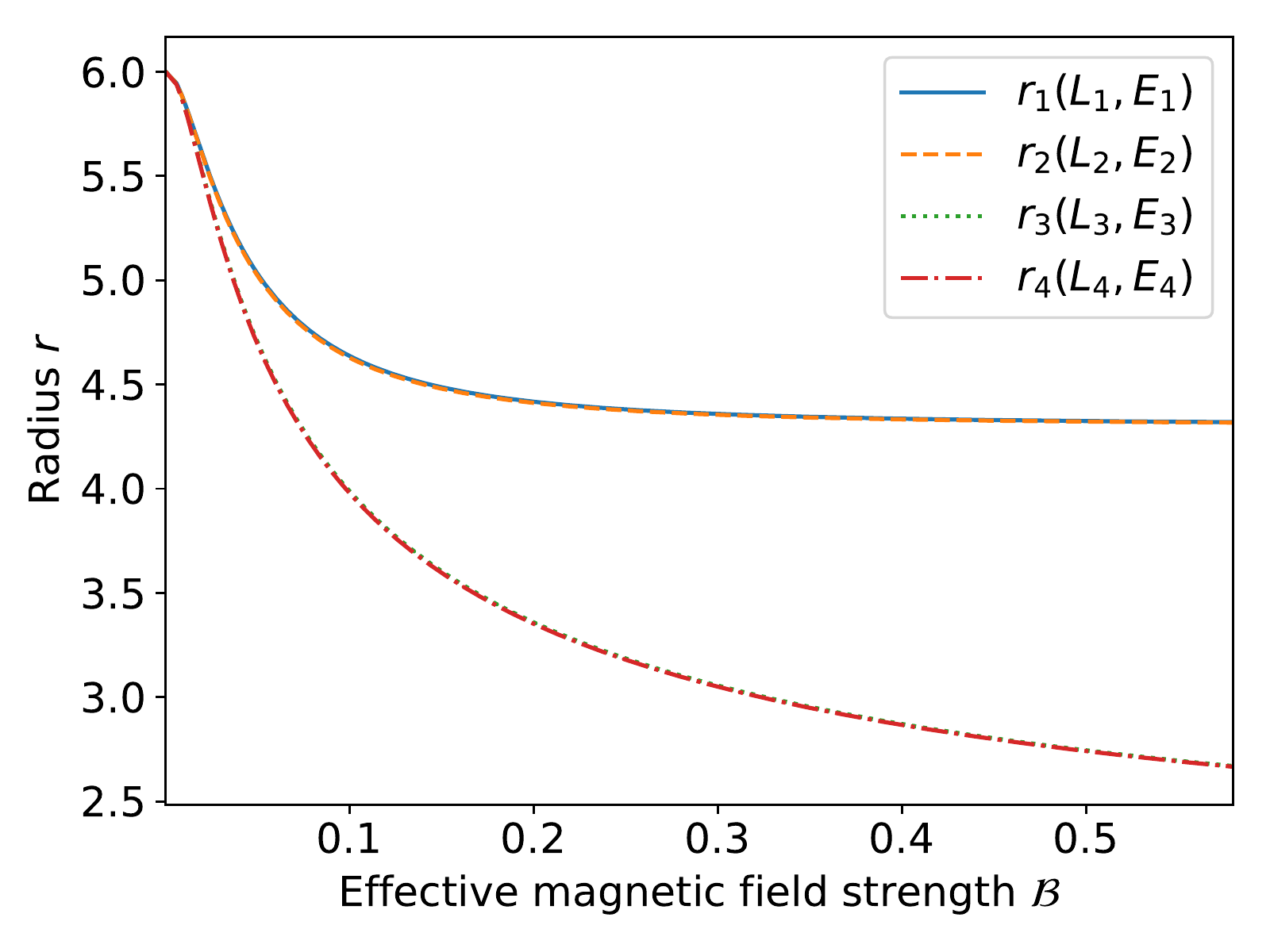}
	\includegraphics[width=0.495\textwidth]{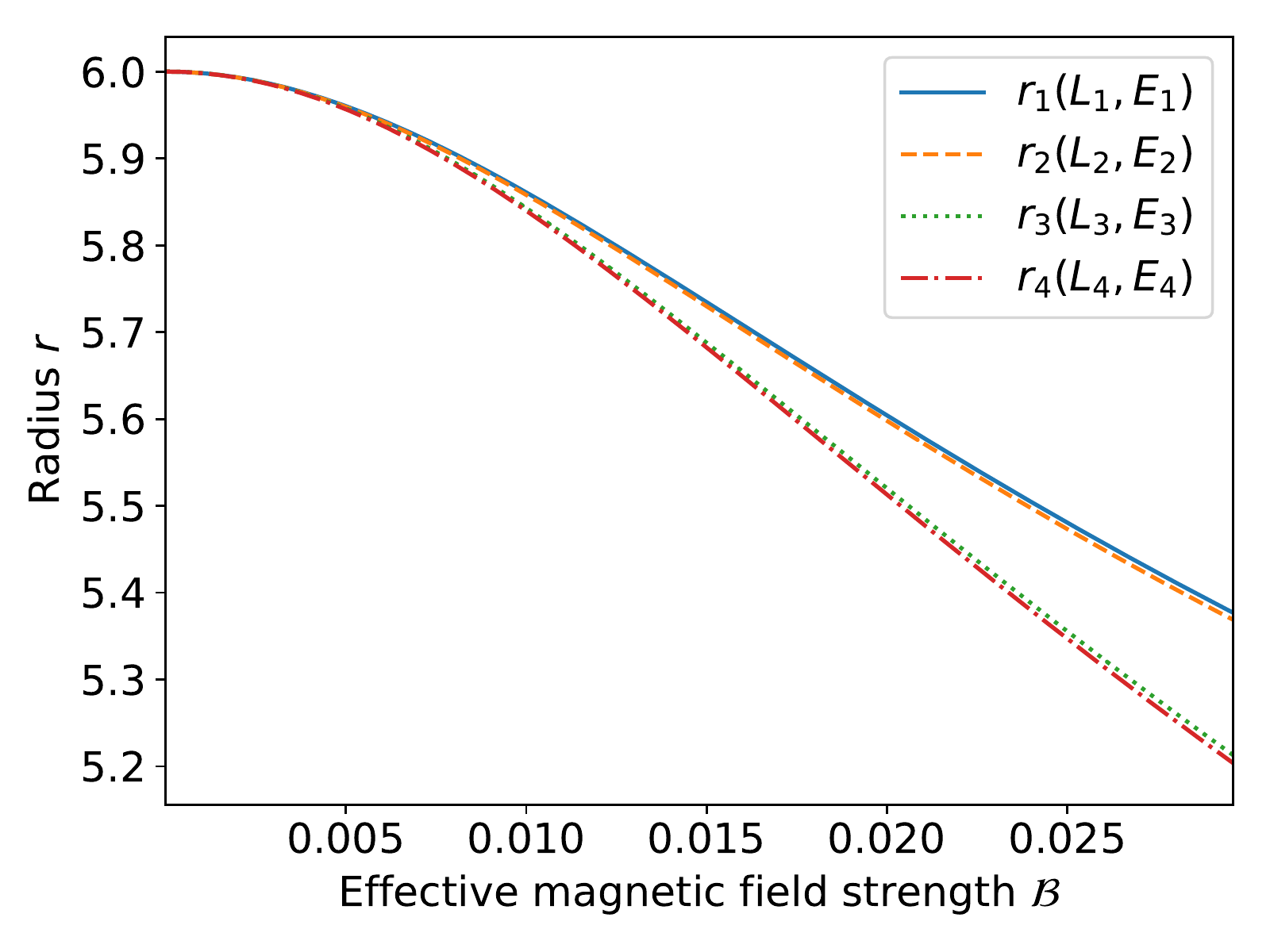}
	\caption{Radius $r_i$ of the ISCO for a charged test particle in an electromagnetic field as a function of the effective magnetic field strength $\mathcal{B}$, where the constant effective electric field strength $\mathcal{Q}$ was chosen in the magnitude of $10^{-3}$. The radii $r_{1,3}$ correspond to orbits of negatively charged particles, $r_{2,4}$ to orbits of positively charged particles. Left plot: General behaviour in the test field. Right plot: Behaviour in the limiting case $\mathcal{B}\rightarrow 0$.}
	\label{fig:1}
\end{figure}

In this case the field's effective electric charge $\mathcal{Q}$ was chosen in the magnitude of $10^{-3}$ for studying infinitesimal effects on the ISCO, while the effective magnetic field strength was defined over the fixed interval $\mathcal{B}=[0,0.6]$. The results for the radial parameter $r$ are illustrated in figure \ref{fig:1}, where the overall behaviour along the fixed interval is shown on the left and $\mathcal{B}\ll 1$ is considered on the right.

While a growing gap between $r_2$ and $r_3$ can be observed for increasing $\mathcal{B}$, the difference between solutions $r_{1,2}$ as well as $r_{3,4}$ are nearly indistinguishable regardless of the magnetic field strength. All four solutions approach (approximately) $r=6$, the ISCO of a neutral test particle, for $\mathcal{B} \approx 0$, agreeing with results from figure \ref{fig:7}. For more details about the case of small $\mathcal{B}$ see the next section.

\begin{figure}
	\centering
	\includegraphics[width=0.495\textwidth]{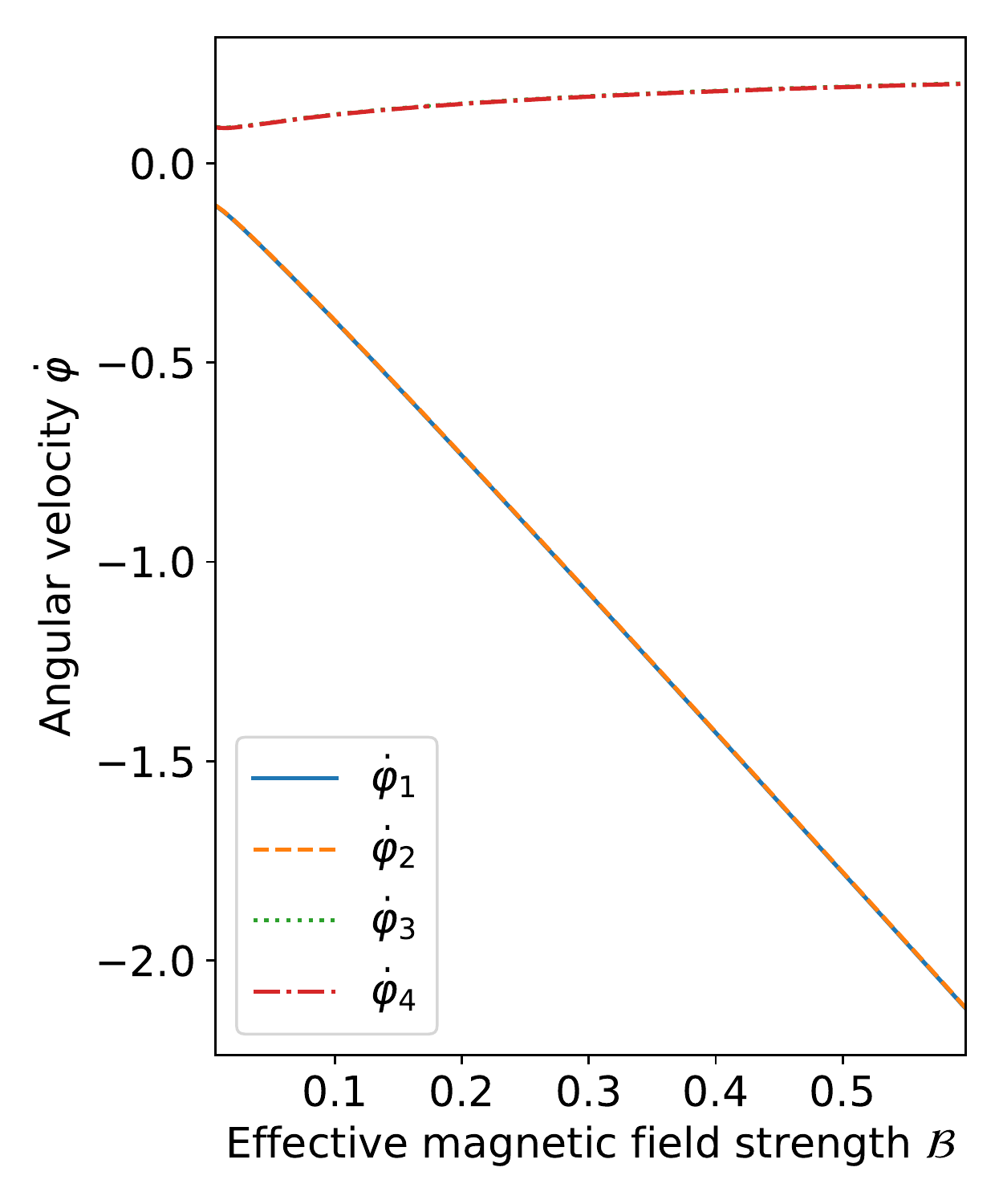}
	\includegraphics[width=0.495\textwidth]{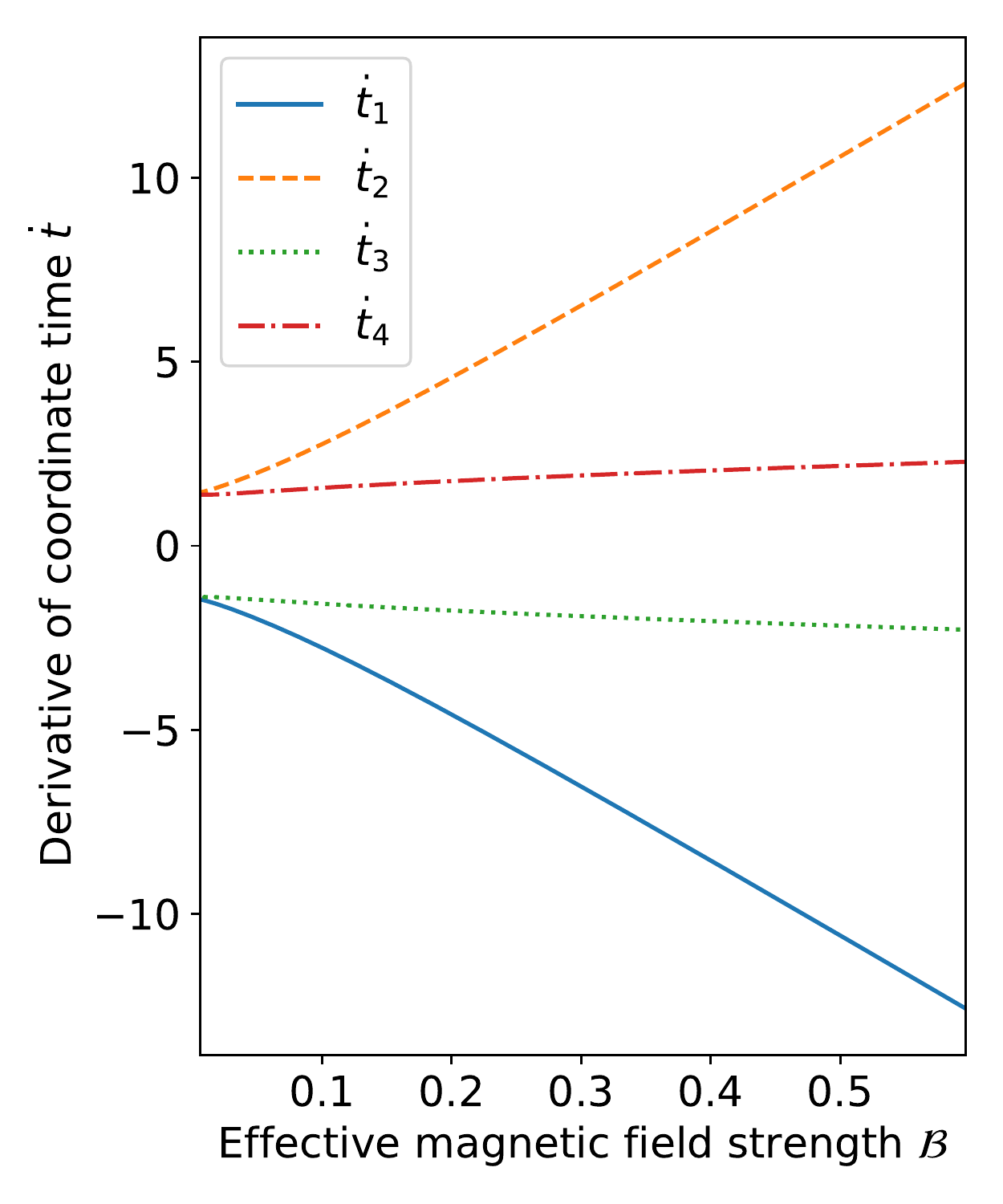}
	\caption{The angular velocity $\dot{\varphi}$ (left plot) and derivative of coordinate time $\dot{t}$ (right plot) of a charged test particle orbiting along the ISCO as a function of the effective magnetic field strength $\mathcal{B}$, where the constant effective electric field strength $\mathcal{Q}$ was chosen in the magnitude of $10^{-3}$.	}
	\label{fig:3}
\end{figure}

Inserting all radii shown in figure \ref{fig:1} into the equations \eqref{Leq} and \eqref{Eeq} individually yields each particle's effective energy $E_i$ and effective angular momentum $L_i$. These in turn can be used to calculate $\dot{t}$ and $\dot{\varphi}$, that are illustrated in figure \ref{fig:3}.

The solutions $r_i$ do not change by simultaneously inverting the signs of $(E,\mathcal{Q})$ or $(L,\mathcal{B})$, as the effective potential is invariant under these transformations. However, $\dot{t}$ and $\dot{\varphi}$, respectively, change their signs. As we chose $B>0$ and $Q>0$ (remember $\mathcal{Q}=qQ/2$, $\mathcal{B}=qB/2$), we simultaneously switch the signs of $\mathcal{Q}$ and $\mathcal{B}$ if we switch the sign of the charge $q$ of the particle. Therefore, we can interpret the particle with ISCO radius $r_1$, that has $\dot{t}<0$ and $\dot{\varphi}<0$ in figure \ref{fig:3}, as a negatively charged particle with $\dot{t}>0$ in a direct orbit $\dot{\varphi}>0$. Both the Lorentz and the Coulomb force consequently point radially inwards, compare also equation \eqref{relF}.

These two attracting forces have to be compensated when forming a stable orbit. Considering figure \ref{fig:1}, the ISCO radius monotonically decreases for increasing $\mathcal{B}$. These results agree with similar observations that have already been made in the case of a magnetic field only \cite{Frolov2010}, even though a physical explanation for this behaviour is not obvious. But $r_1$ being the overall largest solution can still be understood when interpreting $r_2$ to $r_4$ in the same manner and comparing them to $r_1$.

The solution $r_2$, that has $\dot{t}>0$, $\dot{\varphi}<0$ in figure \ref{fig:3}, corresponds to a positively charged particle in an indirect orbit. Because of its positive charge, the Lorentz force still points radially inwards. However in this case, the sign of the particle's charge $q$ coincides with the field's electric charge $Q$, yielding a Coulomb force pointing radially outwards. These opposing forces now cancel each other out to some extent, leading to a lowering of the total force on the particle. This results in a lowering of $r_2$ relative to $r_1$ which is barely visible in figure \ref{fig:1}. Since $Q$ was chosen to be very small, the Lorentz force was predominantly acting on the particle when compared to the Coulomb force. As a consequence the observed difference between $r_{1,2}$ is of order $10^{-3}$.

Analogously, $r_3$ with $\dot{t}<0$ and $\dot{\varphi}>0$ in figure \ref{fig:3} can be assigned to a negatively charged particle in an indirect orbit with $\dot{\varphi}<0$. The Lorentz force points radially outwards, while the Coulomb force attracts the test particle. Since the Lorentz force is much stronger than the Coulomb force for the case $\mathcal{Q}\ll 1$ under discussion, the total electromagnetic force given in  \eqref{relF} will be positive. Compared to $r_2$, the force reverses its orientation, resulting in the considerably larger lowering from $r_1$ to $r_3$ observed in figure \ref{fig:1}. The monotonically decreasing behaviour becomes comprehensible in this case when examining the total electromagnetic force on the orbiting particle. The electromagnetic force is mostly determined by the repelling Lorentz force, leading to the particle being pushed outwards when regarding the ISCO. To create a stable orbit, the radius $r$ decreases, causing stronger gravitation to balance out the electromagnetic force on the particle.

The last possibility is covered by $r_4$, describing a positively charged particle with $\dot{t}>0$ in a direct orbit $\dot{\varphi}>0$. Both Lorentz and Coulomb force point outwards relative to the ISCO, implying even further lowering of $r_4$ compared to the other $r_i$ and thus resulting in the overall smallest solution for $r$.

\subsection{Electromagnetic field with $\mathcal{Q}<\frac{1}{2}$}
\label{Q2}
\begin{figure}
	\centering
	\includegraphics[width=0.495\textwidth]{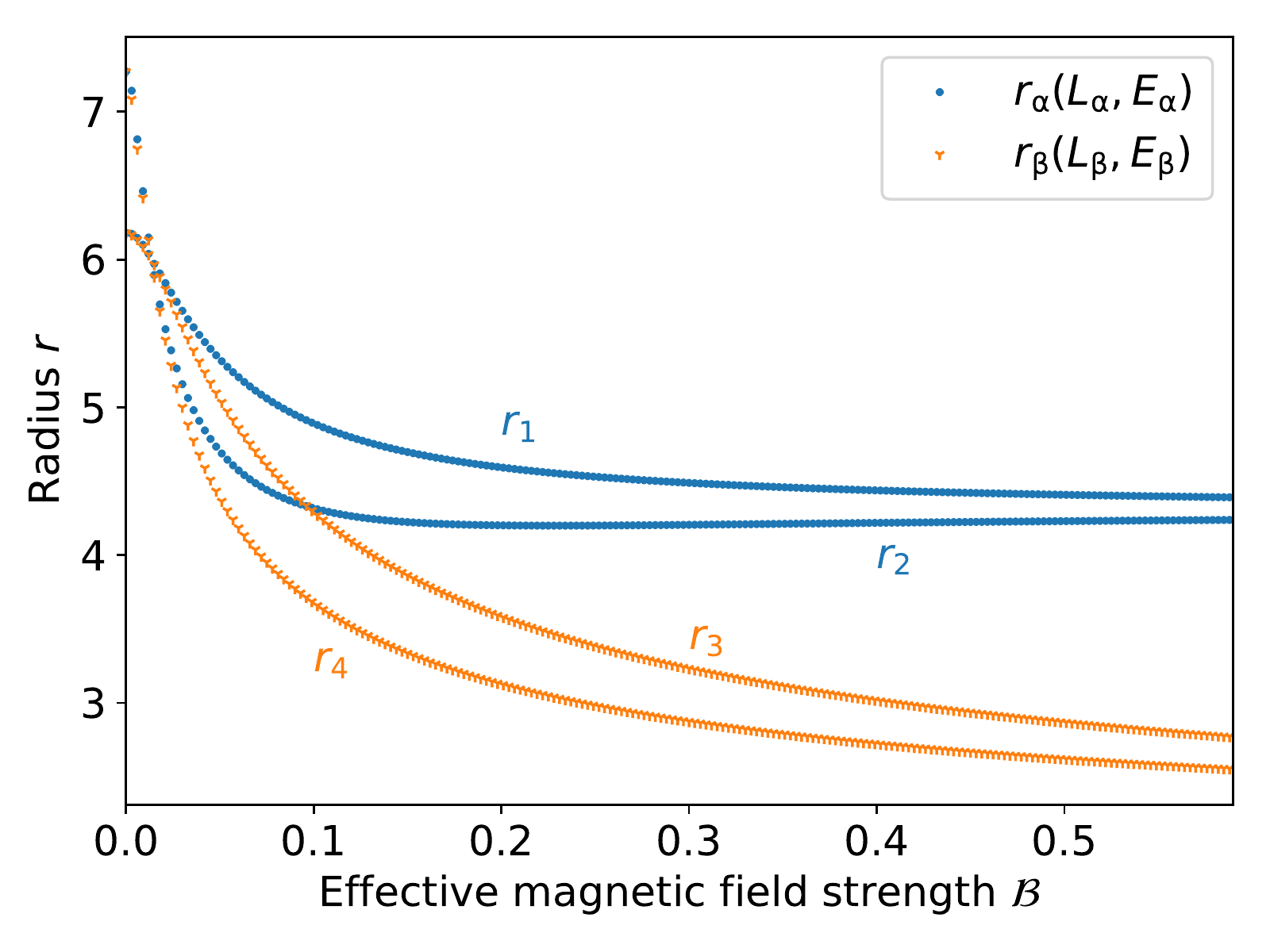}
	\includegraphics[width=0.495\textwidth]{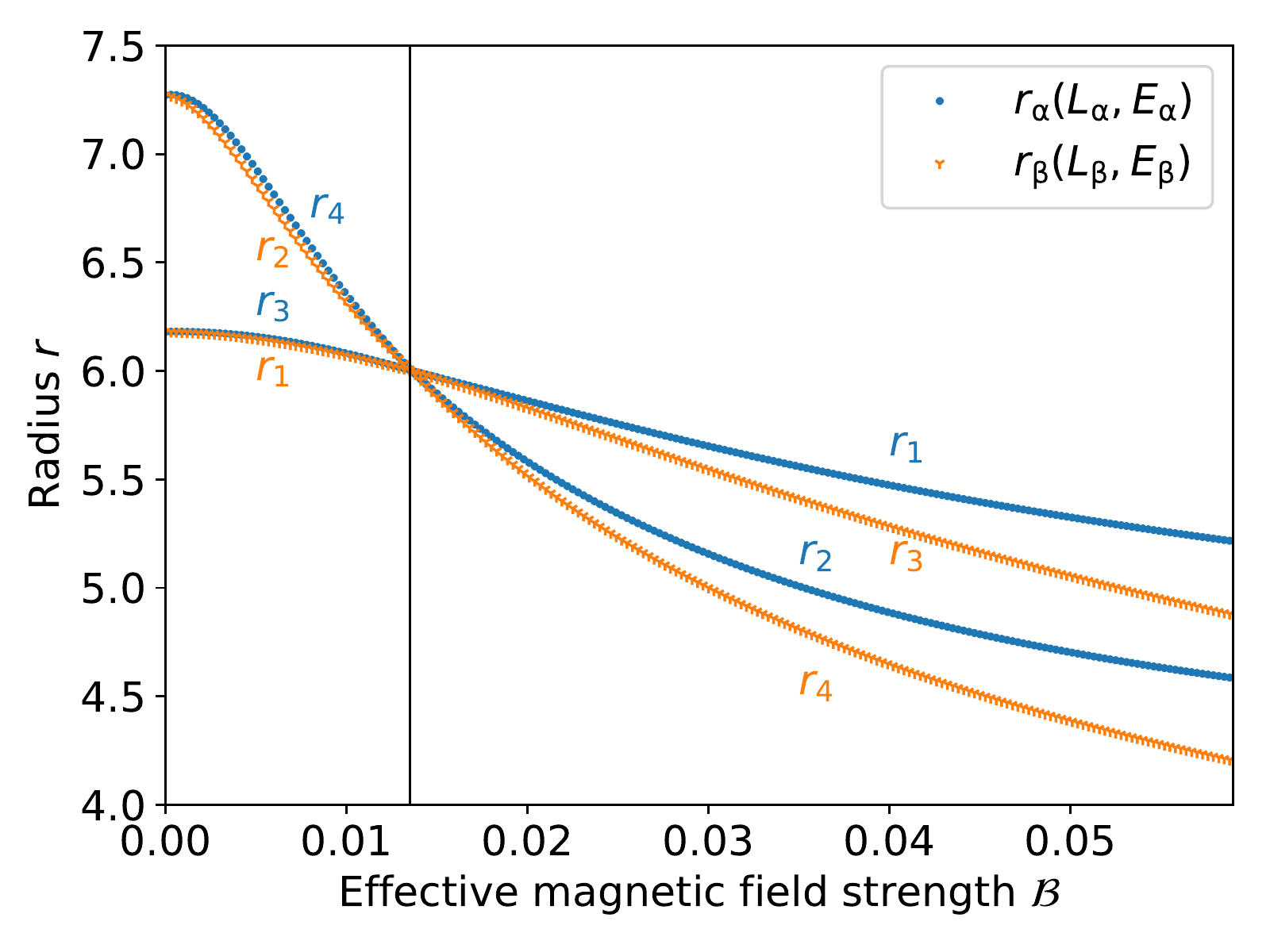}
	\caption{Radius $r_i$ of the ISCO for a charged test particle in an electromagnetic field as a function of the effective magnetic field strength $\mathcal{B}$ with a constant effective electric field strength $\mathcal{Q}=0.4$. Each solution $r_{\upalpha,\upbeta}$ yields two radii $r_i$ corresponding to orbits of either negatively ($r_{1,3}$) or positively ($r_{2,4}$) charged particles. Left plot: General behaviour in the test field. Right plot: Behaviour in the limiting case $\mathcal{B}\rightarrow 0$. The black, vertical line emphasizes all radii intersecting at $r=6$.}
	\label{fig:4}
\end{figure}

Figure \ref{fig:4} depicts the ISCO radius $r$ with increased electric field strength $\mathcal{Q}=0.4$, while the $\mathcal{B}$-interval remained unchanged. Differing from the prior subsection, two intersections are now visible for $\mathcal{B}\neq 0$. Due to the numerical approach of calculating with a certain step size, the second intersection could not be located precisely, while the first was easily found at $r=6$ analytically, coinciding with the ISCO of a test particle in the Schwarzschild metric without external fields.

As will be seen in the next subsection, increasing the electric field strength $\mathcal{Q}$ shifts both intersections to higher magnetic field strengths $\mathcal{B}$. While the second point approaches the black hole horizon in the process, the first intersection remains at constant radius. By inserting $r=6$ into the conditions \eqref{con}, the exactly same analytical expression of the magnetic field strength $\mathcal{B}(\mathcal{Q})$ was derived for all $r_i$, yielding
\begin{equation}
\mathcal{B}=\frac{\sqrt{6}}{72}\mathcal{Q} \quad \text{at }r=6.
\label{rconst}
\end{equation}
Hence, one intersection is to be expected at $r=6$ in the presence of an electromagnetic field, independent of the electric field strength's absolute value. This implies an intersection at $r=6$ to be present for $\mathcal{Q}\ll 1$ in figure \ref{fig:1}. Because $\mathcal{Q}$ was chosen in the magnitude of $10^{-3}$, the corresponding $\mathcal{B}$ was even smaller and in turn not distinguishable from $\mathcal{B}=0$ on the examined interval in figure \ref{fig:1}.

\begin{figure}
	\centering
	\includegraphics[width=0.495\textwidth]{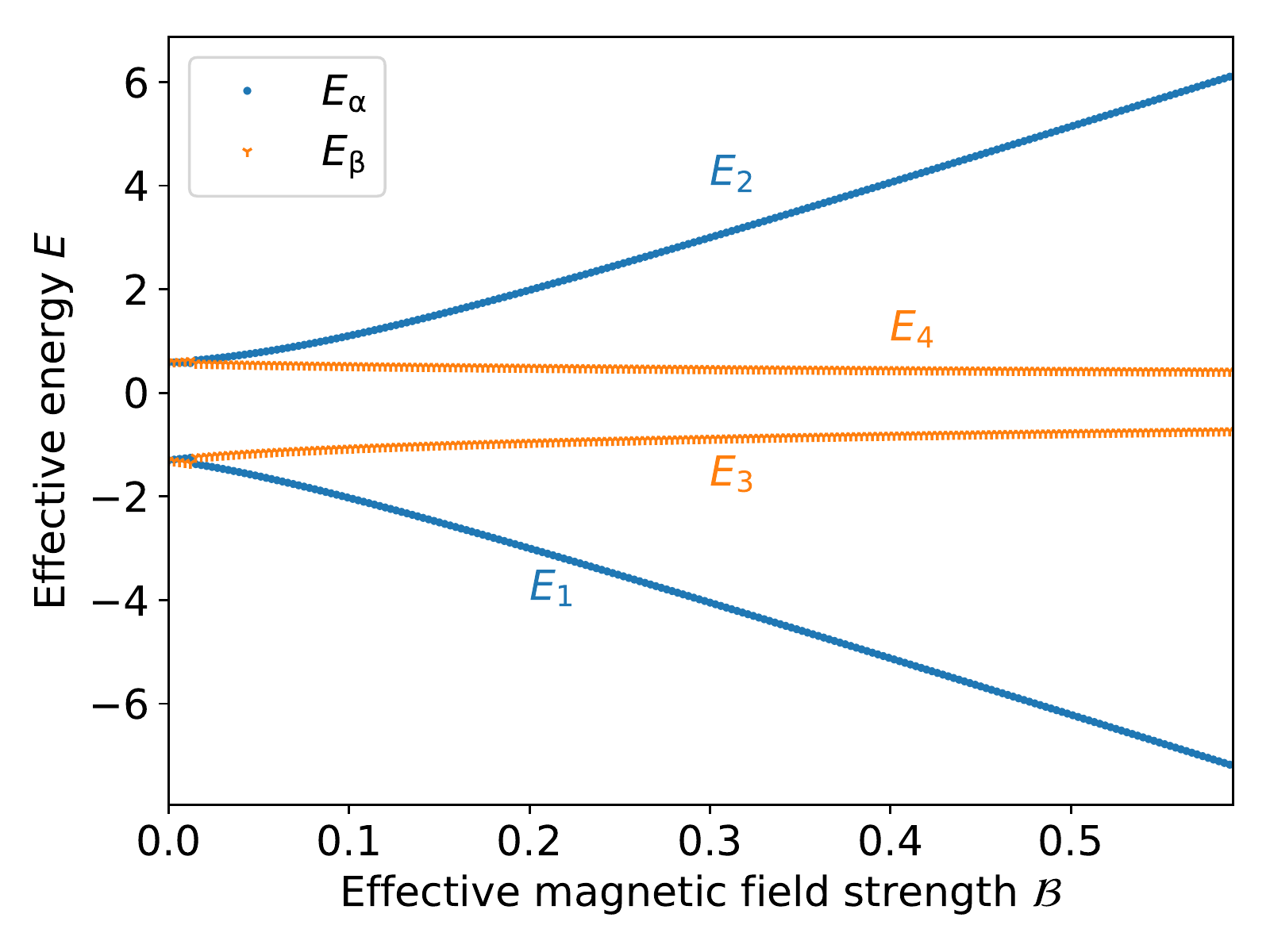}
	\includegraphics[width=0.495\textwidth]{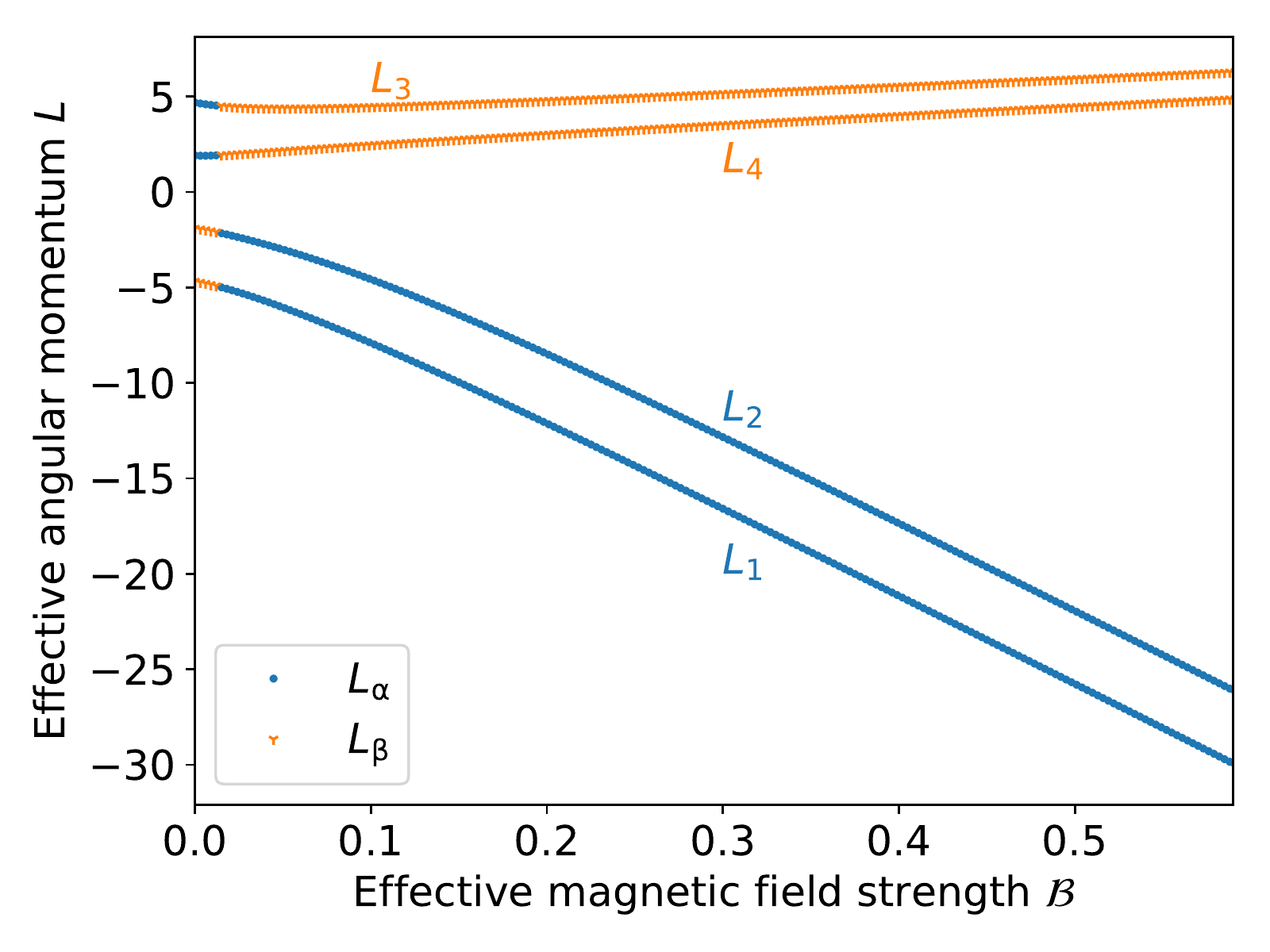}
	\caption{The effective energy $E_i$ (left plot) and effective angular momentum $L_i$ (right plot) of a charged test particle orbiting along the ISCO as a function of the effective magnetic field strength $\mathcal{B}$ with a constant effective electric field strength $\mathcal{Q}=0.4$. The solutions $E_{\upalpha,\upbeta}$ and $L_{\alpha,\beta}$ together show continuous behaviour, resulting in each line corresponding to a particle with a fixed charge and orbiting direction.}
	\label{fig:5}
\end{figure}

The two regions $r>6$ and $r<6$ will be discussed by investigating the behaviour of the electromagnetic force given by equation \eqref{relF}. The graphical representation of the obtained results was changed to allow for a more precise explanation of the actual mathematical solutions and the introduced physical interpretation. The lines were replaced with coloured dots, where each dot corresponds to a calculated value and the colours describe the solutions belonging to $r_\upalpha(L_\upalpha,E_\upalpha)$ or $r_\upbeta(L_\upbeta,E_\upbeta)$, respectively. When examining the angular momenta $L_i$ and the energies $E_i$ calculated from \eqref{Leq} and \eqref{Eeq} in figure \ref{fig:5}, each dotted line corresponds to a particle with properties discussed in subsection \ref{Q1}.

As can be seen, the two solutions $L_{\upalpha,\upbeta}$ as well as $E_{\upalpha,\upbeta}$ yield four continuous lines, respectively. Since it is reasonable to assume a particle's characteristics to behave steadily, each continuous line was assigned to a particle. This was represented by labelling the dotted lines with a corresponding $E_i$ or $L_i$ in figure \ref{fig:5}, such that the order of smallest to highest energy and angular momentum agrees with the results in figure \ref{fig:3}. Each line in figure \ref{fig:4} consequently illustrates the continuous radii $r_i(E_i,L_i)$ analogously to the previous subsection.

\begin{figure}
	\centering
	\includegraphics[width=0.495\textwidth]{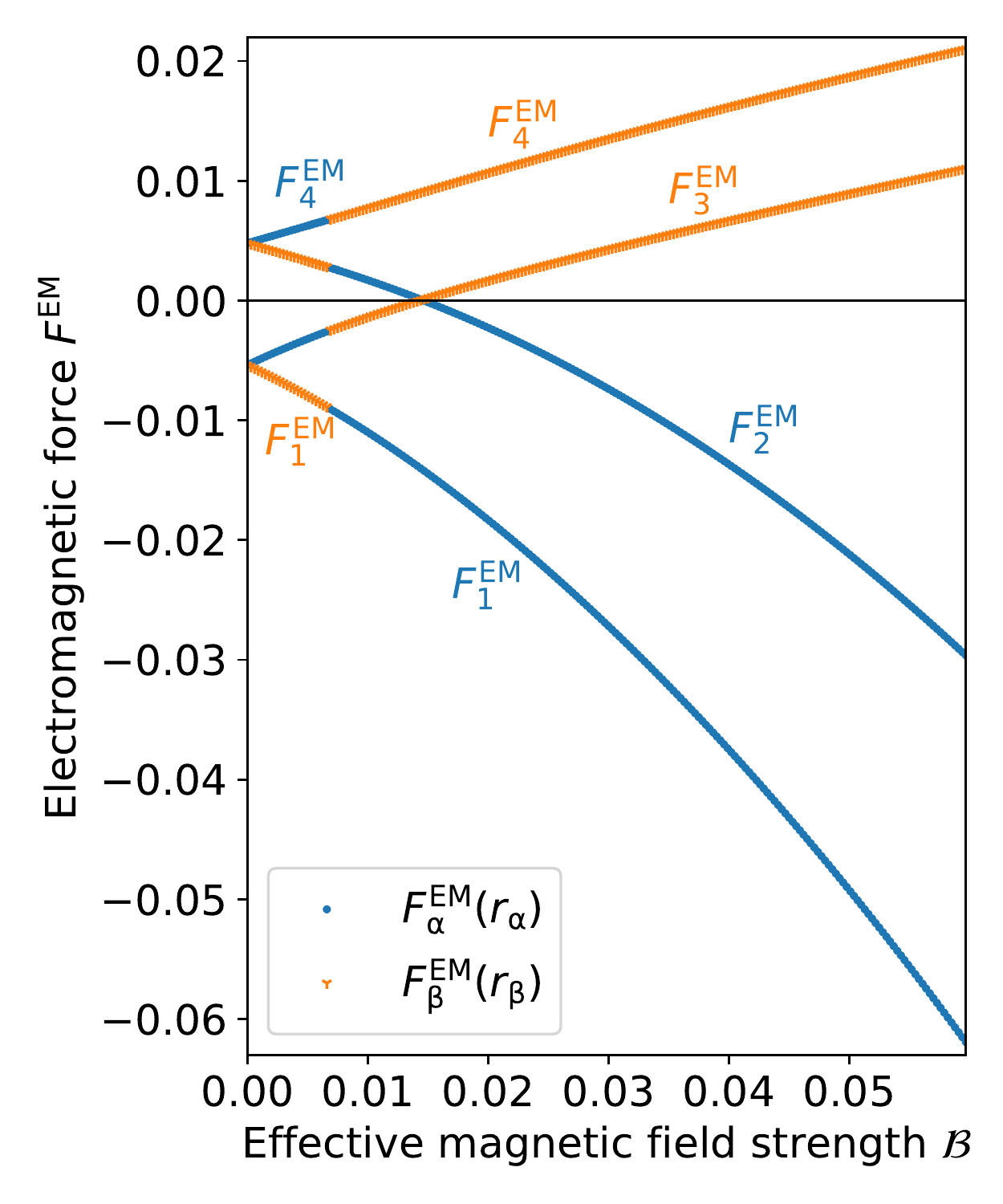}
	\includegraphics[width=0.495\textwidth]{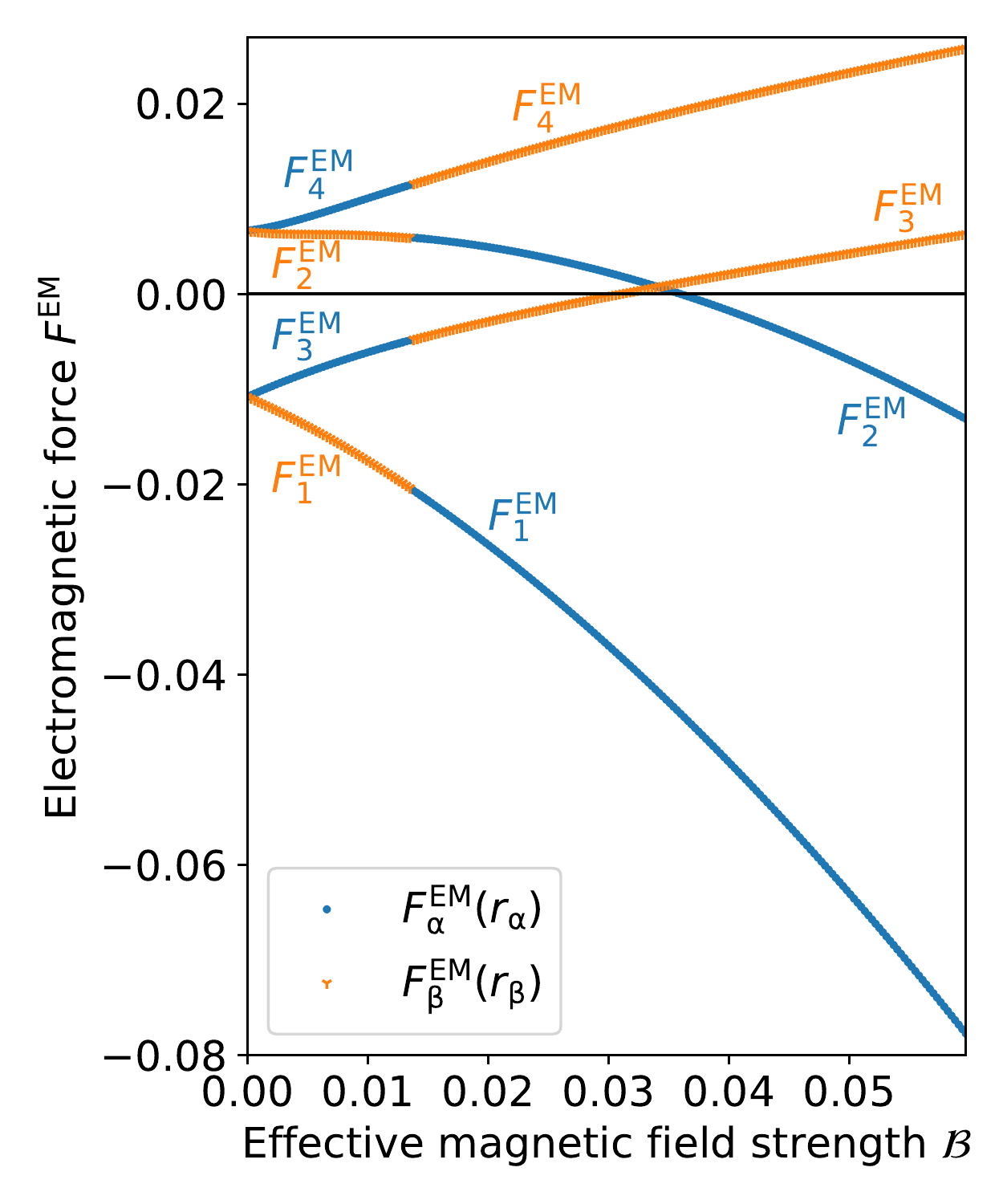}
	\caption{The electromagnetic force $F^\text{EM}_i$ on a charged test particle orbiting along the ISCO in an electromagnetic field as a function of the effective magnetic field strength $\mathcal{B}$ with constant effective electric field strength $\mathcal{Q}=0.2$ (left plot) and $\mathcal{Q}=0.4$ (right plot). The two points $F^\text{EM}_{2,3}=0$ each denote an equilibrium of forces between the opposing Lorentz and Coulomb force, which shift to higher $\mathcal{B}$ for growing $\mathcal{Q}$.}
	\label{fig:9}
\end{figure}

To examine the changes in $r_i$ for an increasing electric field strength, the electromagnetic force $F^\text{EM}_i$ acting on each test particle is illustrated in figure \ref{fig:9} for $\mathcal{Q}=0.2$ and $\mathcal{Q}=0.4$. Both graphs show a behaviour that agrees with the previous subsection. Again, $r_{1,4}$ denote Lorentz and Coulomb force both being attractive or repulsive, respectively. This results in the overall smallest and largest solutions $F^\text{EM}_{1,4}$.  

In comparison, $r_2$ denotes a repulsive Coulomb and an attractive Lorentz force. The Lorentz force vanishes for infinitesimal $\mathcal{B}$, meaning the total force is predominantly determined by the positive Coulomb force. For increasing $\mathcal{B}$, the opposing Lorentz force grows in absolute value until cancelling out the Coulomb force, effectively reaching $F^\text{EM}_{2}=0$ in figure \ref{fig:9}. The attractive Lorentz force becomes larger than the Coulomb force for even higher $\mathcal{B}$, resulting in $F^\text{EM}_{1}<F^\text{EM}_{2}<0$.

Analogously, $r_3$ corresponds to an attractive Coulomb and a repulsive Lorentz force, where $\mathcal{B}\rightarrow0$ yields $F^\text{EM}_{3}<0$. Increasing the magnetic field strength results in the Lorentz force cancelling out the Coulomb force before the total electromagnetic force repels the particle, such that $F^\text{EM}_{4}>F^\text{EM}_{3}>0$. 

For this reason, the radial solutions in figure \ref{fig:4} resemble the results found in figure \ref{fig:1} (with $\mathcal{Q}\ll 1$) for sufficiently large $\mathcal{B}$, but the distance between $r_{1,2}$ as well as $r_{3,4}$ grows with growing $\mathcal{Q}$.

The two values of $\mathcal{B}$ where $F^\text{EM}_{2}$ or $F^\text{EM}_{3}$ cross zero mark a transition of the dominant force on the respective particle from the Coulomb to the Lorentz force as explained above. These transitions happen both in between the two intersection points of the radial ISCO solutions shown in figure \ref{fig:4}. As the order of the ISCO solutions $r_i$ is reversed at the intersection points, we can interpret the whole region from approximately the first intersection at $r=6$ up to about the second intersection as a transitional region.

Concentrating on $r>6$, or equivalently $\mathcal{B}<\frac{\sqrt{6}}{72} \mathcal{Q}$, this results in $r_4>r_2>r_3>r_1$. Here $r_{2,4}$ correspond to positively charged particles and $r_{1,3}$ correspond to negatively charged particles. Between the ISCOs of oppositely charged particles a large radial distance can be observed due to the Coulomb force, whereas the radial distance between the ISCOs of particles with identical sign of the charge is tiny. Note that the $r=6$ intersection corresponds to the change in colour of the continuous lines in figure \ref{fig:9}. For $r<6$, or equivalently $\mathcal{B}>\frac{\sqrt{6}}{72} \mathcal{Q}$, but before the second intersection of ISCO radii, we find $r_1>r_3>r_2>r_4$. Here we cannot clearly identify a dominating Lorentz or Coulomb force. Finally, for large $\mathcal{B}$, after the second intersection of ISCO radii, the Lorentz force dominates, resulting in $r_1>r_2>r_3>r_4$ similar to the case of $\mathcal{Q} \ll 1$ discussed in subsection \ref{Q1}.

In the limiting case $\mathcal{B}=0$, the magnetic field and thus the Lorentz force on the particle vanishes. This in turn reduces the number of solutions for the radius $r$ from four to two, which are caused by the Coulomb force alone and correspond to particles of opposite charge $q$. These two solutions are additionally identifiable in figure \ref{fig:7}. While the respective energies $E_i$ behave accordingly by intersecting in pairs at $\mathcal{B}=0$, four solutions for the angular momentum still exist. Since equation \eqref{rdot} only depends on $L^2$ in the case of $\mathcal{B}=0$, as expected we find $|L_1|=|L_3|$ and $|L_2|=|L_4|$.

Finally, a minimum in $r_2(\mathcal{B})$ is visible when closely examining the results in figure \ref{fig:4}. From a numerical analysis, it seems that this minimum is always present, shifting for smaller $\mathcal{Q}$ to larger $\mathcal{B}$ and approaching $\mathcal{B} \to \infty$ for $\mathcal{Q} \to 0$. This minimum becomes more distinct for larger $\mathcal{Q}$ as described in the following subsection.

\subsection{Electromagnetic field with $\mathcal{Q}\geq\frac{1}{2}$}
\label{Q3}
\begin{figure*}[!ht]
	\centering
	\includegraphics[width=0.495\textwidth]{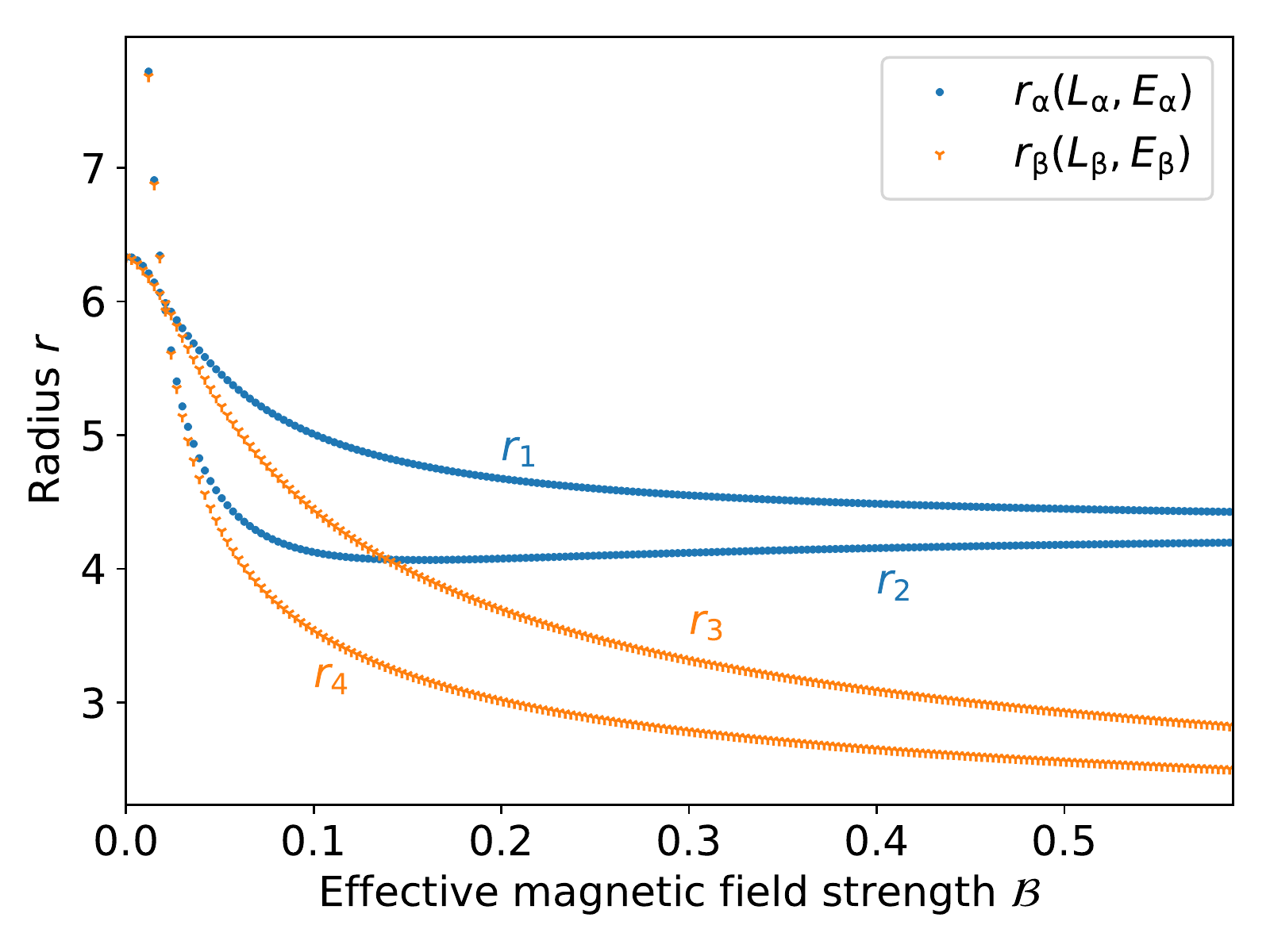}
	\includegraphics[width=0.495\textwidth]{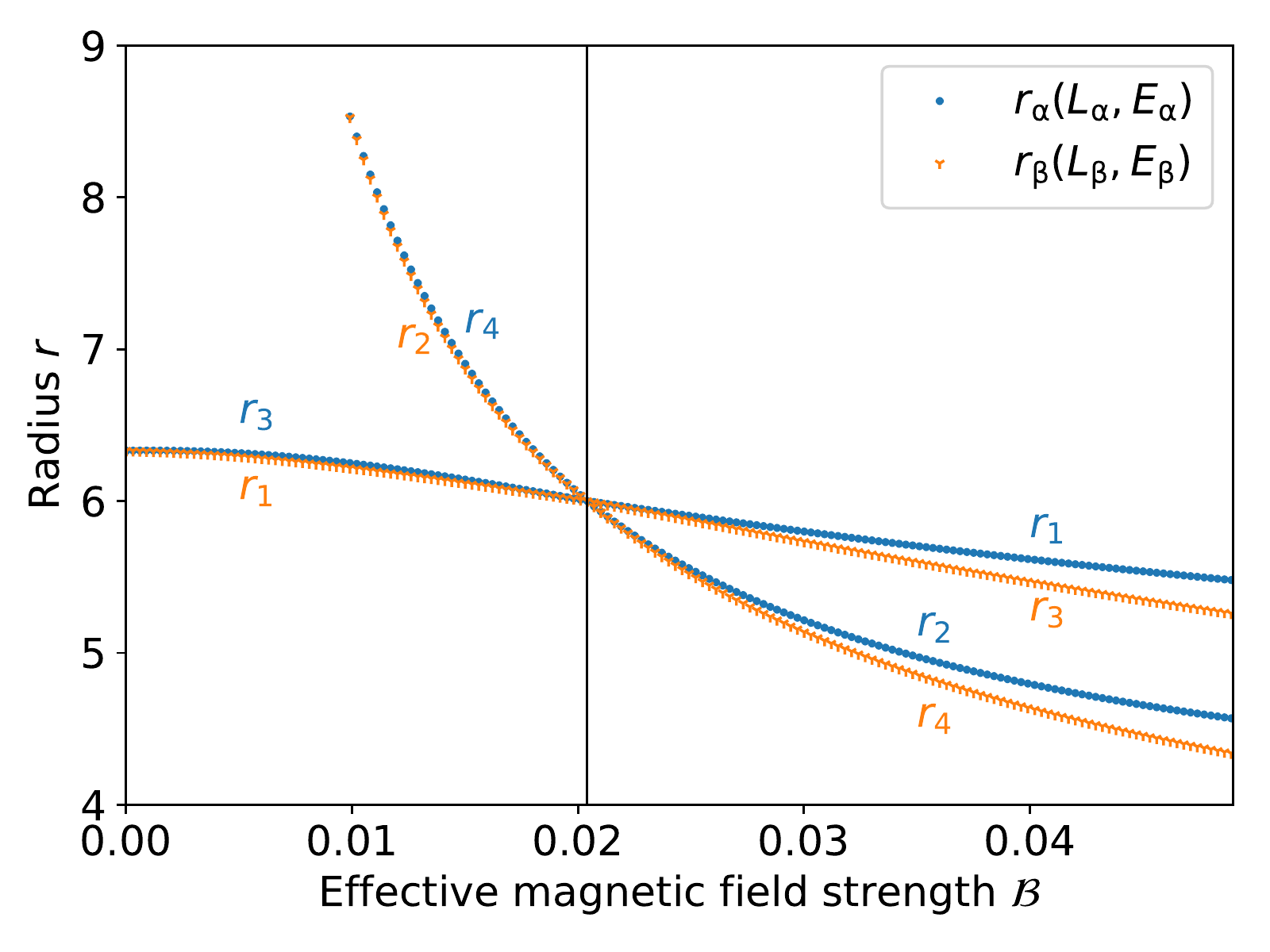}
	\caption{Radius $r_i$ of the ISCO for a charged test particle in an electromagnetic field as a function of the effective magnetic field strength $\mathcal{B}$ with a constant effective electric field strength $\mathcal{Q}=0.6$. Each solution $r_{\upalpha,\upbeta}$ yields two radii $r_i$ corresponding to orbits of either negatively ($r_{1,3}$) or positively ($r_{2,4}$) charged particles. Left plot: General behaviour in the test field. Right plot: Behaviour in the limiting case $\mathcal{B}\rightarrow 0$, where $r_{2,4}$ reach a maximum and then disappear, resulting in a region without ISCOs of positively charged particles. The black, vertical line emphasizes all radii intersecting at $r=6$.}
	\label{fig:6}
\end{figure*}

Further effects can be observed when increasing the electric field strength to $\mathcal{Q}\geq\frac{1}{2}$. The results for the ISCO radius $r$ with an arbitrarily chosen $\mathcal{Q}=0.6$ are illustrated in figure \ref{fig:6}. The four solutions $r_i$ show the expected behaviour in the region $r\leq 6$, or equivalently $\mathcal{B}\geq \frac{\sqrt{6}}{72} \mathcal{Q}$. Equation \eqref{rconst} implies all $r_i$ intersecting at $r=6$ for $\mathcal{B}=\frac{\sqrt{6}}{72} \mathcal{Q} \approx 0.02$, which is slightly larger than for $\mathcal{Q}=0.4$ shown in figure \ref{fig:4}. The intersection of $r_{2,3}$ also shifts to a larger value of $\mathcal{B}$, but additionally lowers in $r$, which can be attributed to the radial shifting of $r_{2}$ and $r_{3}$ due to the electromagnetic force on both particles. The intersections mark a region of balanced electromagnetic forces. An increased $\mathcal{Q}$ induces a greater Coulomb force on the test particle, requiring a stronger Lorentz force to achieve $F^\text{EM}_{2,3}=0$. This in turn shifts both intersections in positive direction along the $\mathcal{B}-$axis in figure \ref{fig:6}.

Considering the case $r>6$, all four $r_i$ behave analogously to the results from figure \ref{fig:4} in the vicinity of $r=6$. For a decreasing magnetic field strength, the distance between the upper and lower $r_i$ pairs increases, while the Lorentz force causes only a tiny difference between the ISCO radii of particles with the same sign of the charge. For $0<\mathcal{Q}<\frac{1}{2}$, the pairs $r_{1,3}$ and $r_{2,4}$ converged at $\mathcal{B}=0$ to the same radius. When the electric field strength reaches $\mathcal{Q}=\frac{1}{2}$, the larger limit value vanishes according to subsection \ref{Qx}. This can be observed on the right of figure \ref{fig:6}, where $r_{2,4}$ first seem to merge to a single radius before they disappear without diverging for sufficiently small $\mathcal{B}\neq 0$.

\begin{figure}
	\centering
	\includegraphics[width=0.495\textwidth]{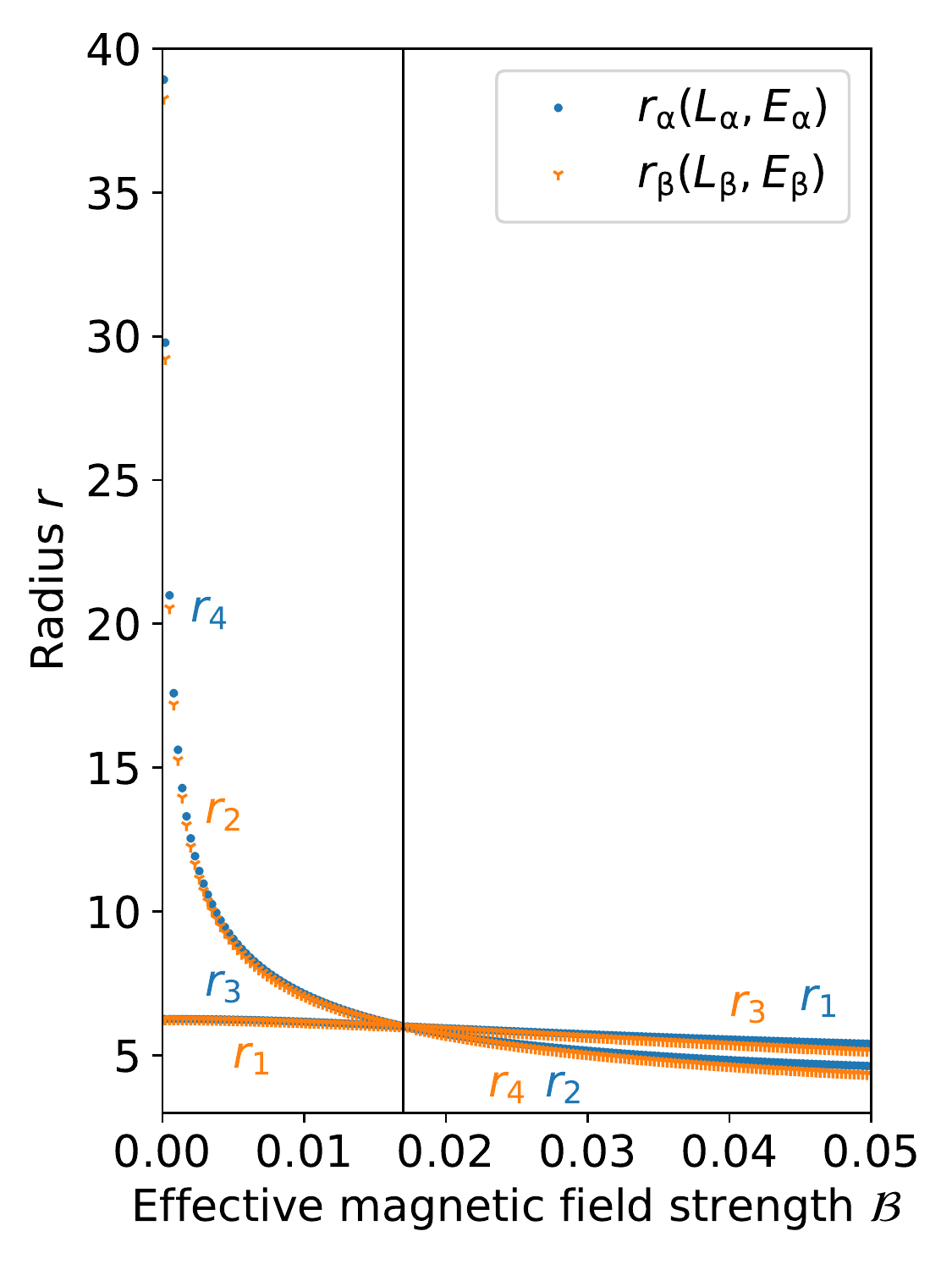}
	\includegraphics[width=0.495\textwidth]{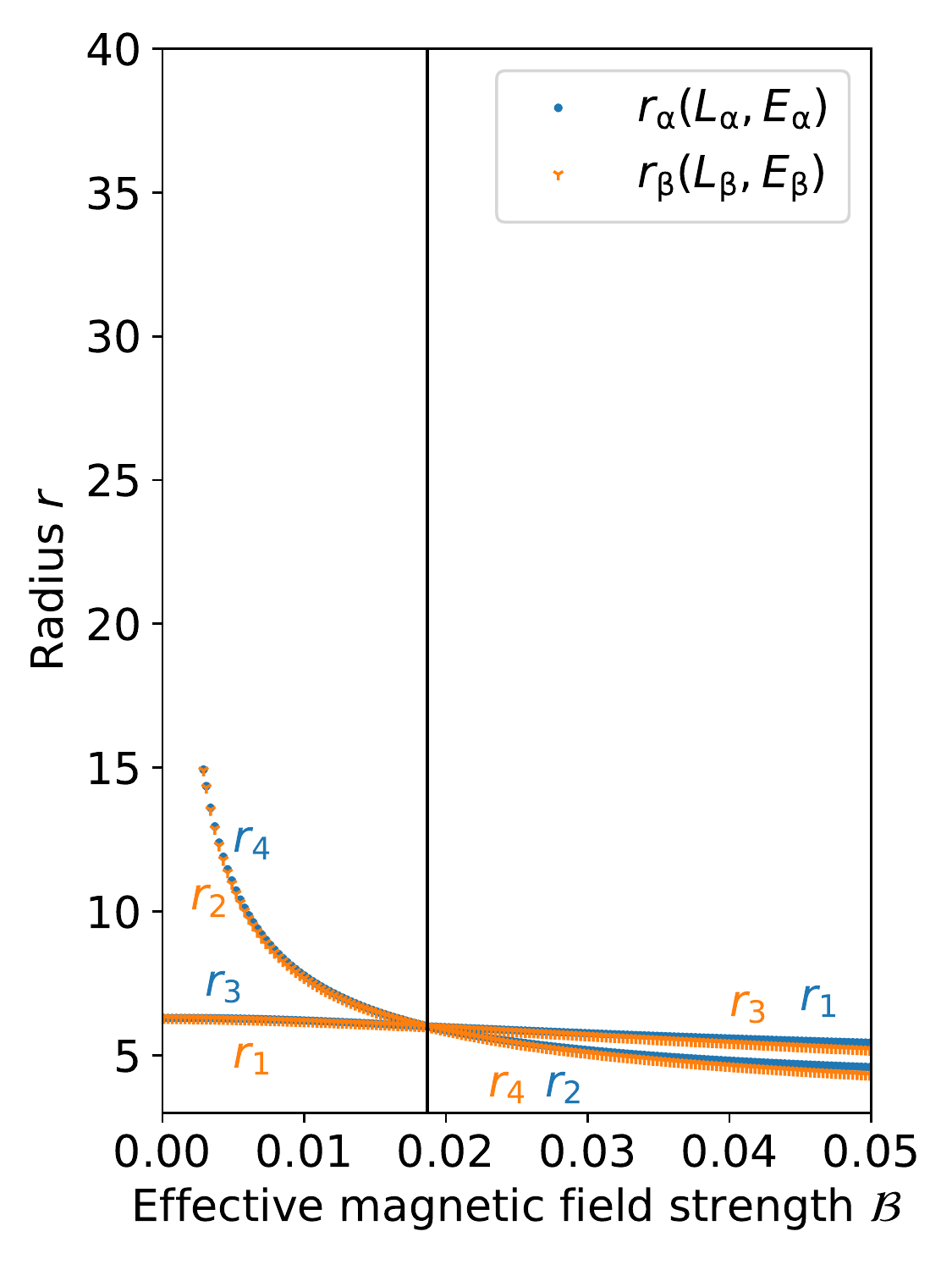}
	\caption{Radius $r_i$ of the ISCO for a charged test particle in an electromagnetic field as a function of the effective magnetic field strength $\mathcal{B}$ with a constant effective electric field strength $\mathcal{Q}=0.5$ (left plot) and $\mathcal{Q}=0.55$ (right plot). Each solution $r_{\upalpha,\upbeta}$ yields two radii $r_i$ corresponding to orbits of either negatively ($r_{1,3}$) or positively ($r_{2,4}$) charged particles. While $r_{2,4}$ diverge for $\mathcal{B}\rightarrow 0$ on the left, they vanish after reaching a maximum for $\mathcal{B}\neq 0$ on the right. This maximum decreases for higher $\mathcal{Q}$. The black, vertical line emphasizes all radii intersecting at $r=6$.}
	\label{fig:8}
\end{figure}

\begin{figure}[ht]
	\centering
	\includegraphics[width=0.495\textwidth]{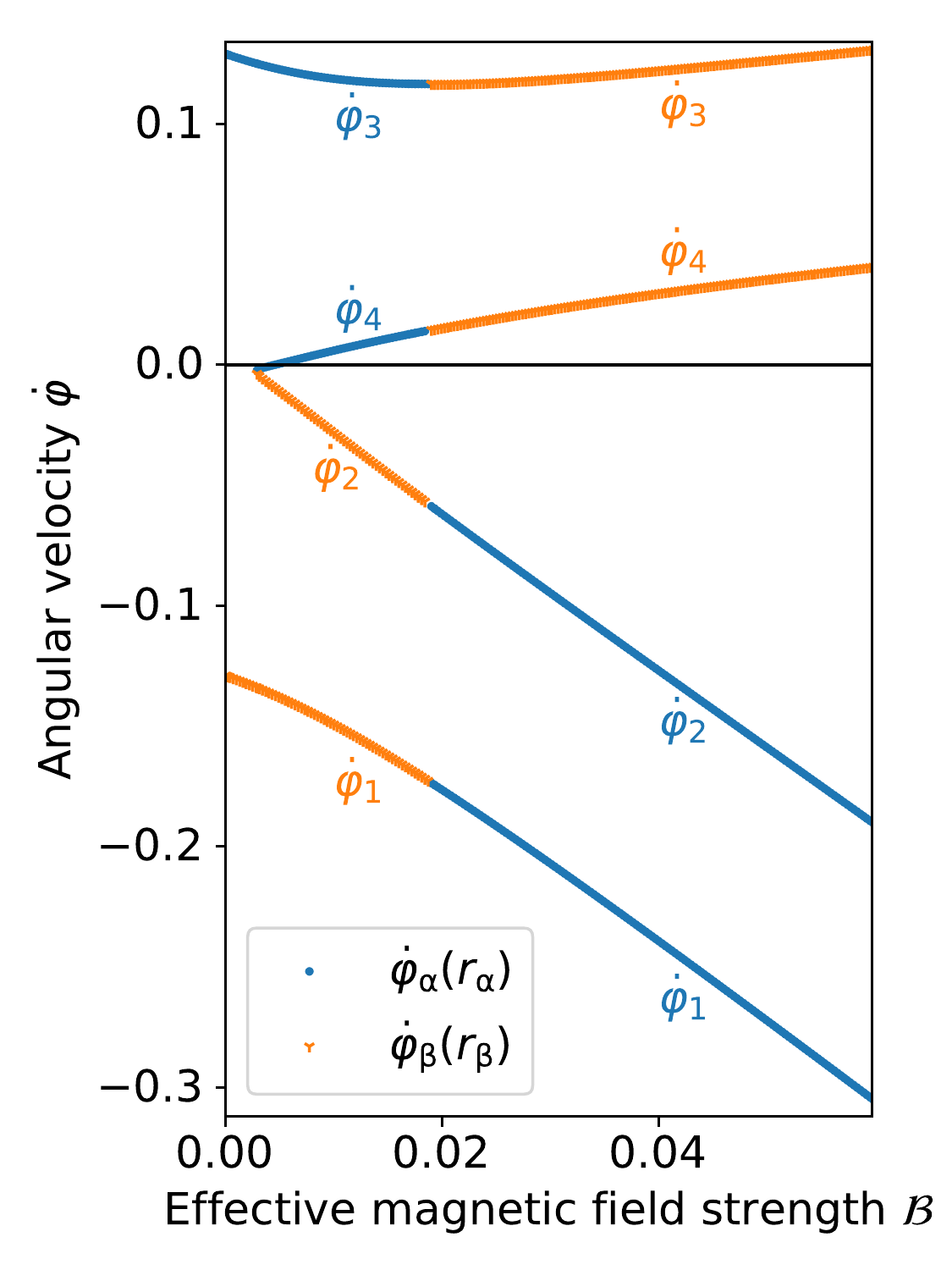}
	\includegraphics[width=0.495\textwidth]{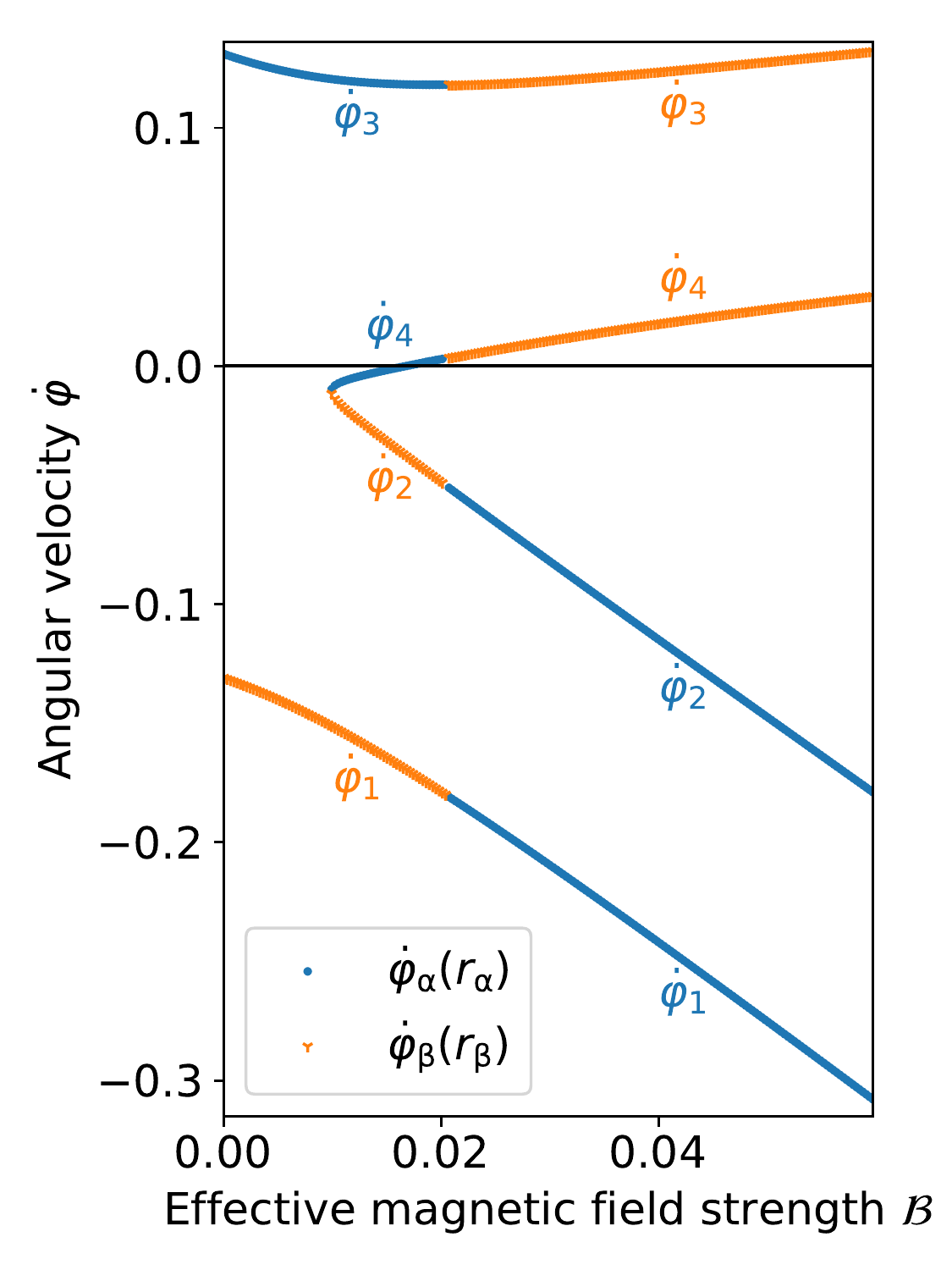}
	\caption{Angular velocity $\dot{\varphi}_i$ of a charged test particle orbiting along the ISCO in an electromagnetic field as a function of the effective magnetic field strength $\mathcal{B}$ with constant effective electric field strength $\mathcal{Q}=0.55$ (left plot) and $\mathcal{Q}=0.6$ (right plot). In both plots, $\dot{\varphi}_4$ vanishes for a certain $\mathcal{B}$, implying a static particle at $r_4$. The point $\dot{\varphi}_4=0$ shifts to higher $\mathcal{B}$ for increased $\mathcal{Q}$.}
	\label{fig:10}
\end{figure}

As the example $\mathcal{Q}=0.6$ illustrates, it is possible to suppress the divergence of the ISCO of positively charged particles for $\mathcal{Q}\geq 0.5$, discussed in subsection \ref{Qx}, by a sufficiently strong magnetic field strength. In the case $\mathcal{Q}=0.6$, the ISCOs with $r_{2,4}$ do not approach from infinity, but rather start from a finite $r-$value that could not be determined precisely. To further examine this behaviour, $\mathcal{Q}=0.5$ and $\mathcal{Q}=0.55$ were chosen and the resulting ISCO radii illustrated in figure \ref{fig:8}.

The electric field strength $\mathcal{Q}=0.5$ characterises the threshold of $r_{2,4}$ vanishing. Calculating the radii in the limiting case $\mathcal{B}=0$ indeed only yields \mbox{ISCOs} $r_{1,3}$ for negatively charged particles. For an infinitesimal $\mathcal{B}\neq 0$ however, all four radii solve the posed conditions. The left graph of figure \ref{fig:8} indicates this with $r_{2,4}$ diverging when approaching $\mathcal{B}=0$. In this case, an infinitesimal magnetic field and thus Lorentz force suffices to allow for the existence of four orbits and results in diverging $r_{2,4}$. Increasing the electric field strength yields a greater Coulomb force, in turn requiring a stronger Lorentz force to permit all four orbits. The right graph of figure \ref{fig:8} shows the resulting finite region where only ISCOs of negatively charged particles exist. After crossing a certain threshold in $\mathcal{B}$, $r_{2,4}$ appear and start to decrease from a maximum. For larger $\mathcal{Q}$ the threshold in $\mathcal{B}$ increases and the maximum of $r_{2,4}$ decreases. The threshold can even expand beyond $\mathcal{B}=\frac{\sqrt{6}}{72}\mathcal{Q}$ for sufficiently large $\mathcal{Q}$, resulting in an intersection of only $r_{1,3}$ at $r=6$.

Finally, a sign change can be observed in $\dot{\varphi}_4$, when plotting the angular velocity $\dot{\varphi}$ for $\mathcal{Q}>\frac{1}{2}$, reversing the particle orbit's direction. This circumstance is illustrated in figure \ref{fig:10} for $\mathcal{Q}=0.55$ and $\mathcal{Q}=0.6$, respectively.

As explained above, the ISCO radii $r_{2,4}$ of positively charged particles only exist for $\mathcal{B}$ larger than a certain threshold value. At this threshold, both positively charged particle ISCOs are indirect orbits with $\dot{\varphi}<0$ as shown in figure \ref{fig:10}, but $L_2<0$ and $L_4>0$. With growing $\mathcal{B}$, $\dot{\varphi}_{2}$ monotonically decreases and $\dot{\varphi}_{4}$ monotonically increases. For a certain finite $\mathcal{B}$, the angular velocity $\dot{\varphi}_{4}$ vanishes, implying the existence of a static particle at fixed angular and radial coordinates. The electromagnetic force in this case is a pure Coulomb force, that is repelling and cancels the gravitational attraction. Note that for slight variations in $\mathcal{B}$ or $\mathcal{Q}$ the particle ISCO will have a very small $\dot{\varphi}$ and, therefore, is stable in this sense. Increasing the electric field strength $\mathcal{Q}$ leads to a smaller $\dot{\varphi}<0$ below the threshold, and $\dot{\varphi}_{4}=0$ additionally shifts to larger $\mathcal{B}$.

Static charged particles have been discussed before in the literature, for instance in a Kerr-Newman spacetime \cite{Balek1989}, and in a Reissner-Nordstr\"om spacetime, see \cite{Pugliese2011,Gladush2009} and references therein. Here, we consider the test field approximation, meaning that this phenomenon is not related to the influence of charge on the curvature of spacetime. Pugliese et al. \cite{Pugliese2011}, considering (electrically) charged particles in a Reissner-Nordstr\"om spacetime with electric charge only, find for black hole spacetimes the same condition on the charge product as we do, namely $\mathcal{Q}>\frac12$. However, in the considered test field approximation, an additional magnetic field of sufficient strength is needed for \textit{stable} circular orbits of positively charged particles, and in particular for the marginally stable static ISCO.

\section{Summary and conclusion}

In this paper we discussed the innermost stable circular orbit (ISCO) of charged particles in a Schwarzschild spacetime endowed with electromagnetic test fields that do not influence the curvature of spacetime. For the electromagnetic fields we chose the Wald solution of an asymptotically uniform magnetic field \cite{Wald74} and a radial electric field centered at the black hole. As the particle motion will in general be chaotic, we restricted our considerations to the equatorial plane, meaning orthogonal to the magnetic field, to ensure integrability.

Due to the four combinations of equal/opposite charges of particle and black hole as well as aligned/anti-aligned angular momentum and magnetic field strength in general four different ISCOs will arise. To simplify the discussion and representation of our results, we chose without loss of generality $Q>0$ for the electric field and $B>0$ for the magnetic field. All other cases can be reconstructed from this by symmetry arguments. Moreover, we chose w.l.o.g. $qQ/2 =: \mathcal{Q}>0$ and $qB/2=:\mathcal{B}>0$, where $q$ is the specific charge of the particle. As the equations of motion are invariant under the transformation $(L,\mathcal{B}) \to (-L,-\mathcal{B})$ as well as $(E,\mathcal{Q}) \to (-E,-\mathcal{Q})$, where $E$ and $L$ are the energy and angular momentum, all other cases can be reconstructed from this.

As the limit of $\mathcal{Q}=0$ was already discussed in detail by Frolov and Shoom \cite{Frolov2010}, we started our analysis of the ISCO with the limit $\mathcal{B}=0$. Similar to earlier results in more general setups, see for instance \cite{Schroven2017}, we found that the ISCO radius of charged particles always increases as compared to neutral particles regardless of the sign of the charge. Moreover, for positively charged particles the ISCO diverges to infinity at $\mathcal{Q}=\frac12$ and vanishes completely for larger $\mathcal{Q}$, compare e.g.~\cite{Pugliese2011}. We then proceeded with the case of $\mathcal{Q}\ll 1$. As expected we found four ISCO solutions for the four different combinations discussed above. However, due to the smallness of the chosen $\mathcal{Q}$, two pairs can be identified that behave almost identical and very similar to the results already discussed by Frolov and Shoom \cite{Frolov2010}. Here we also discussed in some detail the physical interpretation of the mathematical results as a preparation for the following more complicated setups.

As $\mathcal{Q}=\frac12$ clearly presents an interesting limiting case, we split our discussion accordingly, starting with $\mathcal{Q}<\frac12$. We found that all four ISCO solutions will always cross at $r=6$, the ISCO of neutral particles, for $\mathcal{B} = \frac{\sqrt{6}}{72} \mathcal{Q}$. A second intersection of only two of the four solutions can be identified at a larger $\mathcal{B}$, compare figure \ref{fig:4}, whose exact value we could not determine analytically. At each intersection, the ISCO radii reverse their order. In between these two intersections, we found that the electromagnetic force will vanish for two of the four ISCOs, see figure \ref{fig:9}. We therefore conclude that this region of intersections is a transitional region, where neither Coulomb nor Lorentz force clearly dominate. For smaller $\mathcal{B}$ the Coulomb force dominates, with results resembling the $\mathcal{B}=0$ limit, whereas for large $\mathcal{B}$ the Lorentz force dominates, with results resembling the $\mathcal{Q}=0$ limit.

Finally, for $\mathcal{Q}>\frac12$, we found in addition to the characteristics already discussed for the $\mathcal{Q}<\frac12$ case two interesting new features. Firstly, due to the observed nonexistence of an ISCO for positively charged particles in the $\mathcal{B}=0$ limit, we find a region close to $\mathcal{B}=0$ where only two ISCOs (for negatively charged particles) exist. Interestingly, the ISCO for positively charged particles reappears at a finite radial position for sufficiently strong magnetic fields. At exactly $\mathcal{Q}=\frac12$, an infinitesimal magnetic field is sufficient to allow for four ISCOs, but the ISCOs of positively charged particles diverge for $\mathcal{B} \to 0$. Secondly, we found that both ISCOs for positively charged particles reappear initially as indirect orbits with $\dot{\varphi}<0$. However, for one of the particles $\dot{\varphi}$ will increase with increasing $\mathcal{B}$ and eventually cross $\dot{\varphi}=0$, allowing for an ISCO given by a static particle sitting at a fixed radial and angular coordinate. 

The results of this paper show that the structure of stable circular orbits is very rich once electromagnetic (test) fields and charged particles are relevant. As argued in the introduction, electromagnetic test fields can astrophysically be relevant in particular for the motion   of free electrons and protons, that have a very high charge to mass ratio. We therefore think that further research to understand the structure of ISCOs of charged particles, and in particular the fact that the discussed electric test field always increases, whereas the chosen magnetic field always decreases the ISCO radius, is very worthwhile.


\begin{acknowledgements}
The authors thank Kris Schroven for fruitful discussions. We gratefully acknowledge support from the Research Training Group RTG 1620 ``Models of Gravity'' funded by the Deutsche Forschungsgemeinschaft (DFG). E.H. is thankful for support from the DFG funded Cluster of Excellence EXC 2123 ``Quantum Frontiers''.
\end{acknowledgements}

\bibliographystyle{spphys}       
\bibliography{Jan}   

%


\end{document}